\documentclass[linenumbers,fleqn,usenatbib]{mnras}

\usepackage{amsmath,amssymb}

\usepackage{xcolor}

\usepackage{graphicx}

\usepackage{multirow}
\definecolor{purple2}{rgb}{0.5, 0.0, 0.5}
\definecolor{red2}{rgb}{0.9, 0.0, 0.1}

\newcommand{\mrm}[1]{\mathrm{#1}}
\newcommand{\nuc}[2]{$\mrm{^{#2}#1}$}

\title[Nova Gamma Rays]{Gamma-Ray Light Curves and Spectra of Classical Novae}

\author[Leung and Siegert]{
Shing-Chi Leung,$^{1}$\thanks{E-mail: scleung@caltech.edu}
Thomas Siegert,$^{2,3}$
\\
$^{1}$TAPIR, Mailcode 350-17, California Institute of Technology,
Pasadena, CA 91125, USA\\
$^{2}$Institut f\"ur Theoretische Physik und Astrophysik, Universit\"at W\"urzburg, Campus Hubland Nord, Emil-Fischer-Str. 31, \\
97074 W\"urzburg, Germany\\
$^{3}$Max-Planck-Institute for extraterrestrial Physics, Gie\ss enbachstra\ss e 1, 85748, Garching bei M\"unchen, Germany
}

\date{Accepted XXX. Received YYY; in original form 10 Dec 2021}
\pubyear{2021}

\date{\today}

\begin{document}
\label{firstpage}
\pagerange{\pageref{firstpage}--\pageref{lastpage}}
\maketitle

\begin{abstract}
The nucleosynthesis in classical novae, in particular that of radioactive isotopes, is directly measurable by its $\gamma$-ray signature.
Despite decades of observations, MeV $\gamma$-rays from novae have never been detected -- neither individually at the time of the explosion, nor as a result of radioactive decay, nor the diffuse Galactic emission from the nova population.
Thanks to recent developments in modeling of instrumental background for MeV telescopes such as INTEGRAL/SPI and Fermi/GBM, the prospects to finally detect these elusive transients are greatly enhanced.
This demands for updated and refined models of $\gamma$-ray spectra and light curves of classical novae.
In this work, we develop numerical models of nova explosions using sub- and near-Chandrasekhar CO white dwarfs as the progenitor.
We study the parameter dependence of the explosions, their thermodynamics and energetics, as well as their chemical abundance patterns.
%
%
We use a Monte-Carlo radiative transfer code to compute $\gamma$-ray light curves and spectra, with a focus on the early time evolution.
We compare our results to previous studies and find that the expected 511-keV-line flash at the time of the explosion is heavily suppressed, showing a maximum flux of only $10^{-9}\,\mrm{ph\,cm^{-2}\,s^{-1}}$ and thus making it at least one million times fainter than estimated before.
This finding would render it impossible for current MeV instruments to detect novae within the first day after the outburst.
Nevertheless, our time-resolved spectra can be used for retrospective analyses of archival data, thereby improving the sensitivity of the instruments.
\end{abstract}



\section{Introduction}
\label{sec:intro}
\subsection{Nova physics}\label{sec:nova_physics}
Classical novae are transient events originating from the thermonuclear runaway (TNR) on the surface of white dwarfs (WDs), resulting in a mass outburst \citep[see, e.g., reviews from][]{Starrfield2016, Chomiuk2020}.
Multiple physical models exist to explain the diversity of the outburst phenomenon, including mass transfer from a companion star, symbiotic star, and dwarf novae \citep{Webbink1987}.
The mass transfer model has been studied widely:
It includes a binary system in which a compact object, i.e. a WD or neutron star (NS), accretes H- or He-rich matter from its companion main-sequence star through a filled Roche Lobe \citep{Paczynski1965, Starrfield1987}.
The later TNR is developed due to the thermally unstable H-burning shell \citep{Rose1968, Starrfield1971}.
After the mass outburst, the luminosity decreases until it re-approaches the initial luminosity so that the mass accretion can resume \citep{MacDonald1985}.

The dynamics of novae are non-trivial because it might involve the accretion disk with an aspherical outburst.
The ignition and the begin of the mass outburst begin depend on the mass transfer rate from the companion star, the progenitor mass, the composition and mixing \citep{Kovetz1985, Truran1986, Shen2009, Townsley2004, Denissenkov2013, Ginzburg2021}.
The mixing is essential for explaining the observed metal abundances in the ejecta \citep{Prialnik1986}.
The rapidly rotating accretion disk creates a rotational instability that triggers element mixing of the H-rich matter in the accretion disk and the CO- (or ONe-) rich matter of the WD \citep{Prialnik1984,Fujimoto1988}, by turbulent mixing 
\citep{Fujimoto1993}, convective overshooting \citep{Glasner1995} and Kelvin-Helmholtz instability \citep{Jose2020}.
Angular momentum is also transported in the process \citep{Siegfried1989}.
The mixing process driven by the wave breaking is important to the ignition because otherwise, the CO-rich matter cannot diffuse efficiently into the H-envelope, making the TNR weak \citep{Alexakis2004}.

Although novae show an occurrence rate about 100--1000 times higher than supernovae in a galaxy, their contribution of global metal enrichment is small compared to supernovae because they eject a much lower mass ($\lesssim 10^{-7}-10^{-5}\,M_{\odot}$) during the outburst \citep{Hernanz1996, Jose2020}.
Nevertheless, novae are ideal laboratories for the production of low-mass elements that eventually mix with the interstellar medium \citep{Gehrz1998}.
For example, they produce an enhanced abundance in $\beta$-decay isotopes of the CNO cycle including $^{13}$C, $^{15}$N and $^{17}$O \citep{Starrfield1972}.
Most reactions rely on the hot-CNO cycle and proton capture.
The TNR can also trigger formation of intermediate ($^{22}$Na) and longer ($^{26}$Al) lifetime radioactive isotopes \citep{Weiss1990}.
The radioactive buildup from the population of novae in the Milky Way is expected to contribute to the diffuse $\gamma$-ray emission along the Galactic plane \citep{Diehl2021}.
It is also a rich source of $^{7}$Li \citep{Starrfield1978, Starrfield2020}, suggested by the decay signal of $^{7}$Be \citep{Harris1991, Tajitsu2015}.
Directly measuring the yields in nova explosions can help to constrain the uncertainties of nuclear reaction rates \citep{Jose2001} and the mystery of the cosmic origin of lithium and fluorine \citep{Spitoni2018, Ryde2020}.

\subsection{Gamma-ray observations}\label{sec:gamma-ray_obs}
\citet{Clayton1974} pointed out that suited $\gamma$-ray telescopes could observe the essential information to prove the intricacies of the nova phenomenon.
In particular, the authors suggested the 2312\,keV and 1275\,keV line from \nuc{O}{14} and \nuc{Na}{22}, respectively, and the 511\,keV line from various $\beta^+$-unstable isotopes including \nuc{N}{13}, \nuc{O}{14}, and \nuc{O}{15}.
Follow-up studies \citep[e.g.,][]{Clayton1981_novae7Li,Leising1987_nova_511,Hernanz2006,Hernanz2014_nova} then also suggested that the 478\,keV line from \nuc{Be}{7} should be visible up to distances of $\sim 2$\,kpc.
To this day, no individual nova nor the diffuse radioactive glow of the nova population have been detected in soft $\gamma$-rays.

The three major objectives, the 1275\,keV line from \nuc{Na}{22}, the 478\,keV line from \nuc{Be}{7}, and the 511\,keV line flash from various short-lived isotopes, had been studied in several cases:
Using CGRO/COMPTEL, \citet{Iyudin1999_NovaCygni1992} found a limiting ejecta mass of \nuc{Na}{22} of $<2.1 \times 10^{-8}\,\mrm{M_{\odot}}$ from Nova Cygni 1992.
Trying to observe the Galactic ridge, \citet{Jean2001_ONenovae1275} constrained the \nuc{Na}{22} ejecta mass to $<3 \times 10^{-7}\,\mrm{M_{\odot}}$.
By applying a Bayesian hierarchical model to INTEGRAL/SPI data, \citet{Siegert2021_BHMnovae} could limit the \nuc{Na}{22} ejecta mass in the Milky Way to $<2 \times 10^{-7}\,\mrm{M_{\odot}}$ per nova.
The best target for detecting the 478\,keV line within the last 20 years was V5668\,Sgr, because of its proximity of 1--2\,kpc.
\citet{Molaro2016_V5668} detected \nuc{Be}{7}\,II lines in the UV which suggest a \nuc{Be}{7} ejecta mass of $\sim 0.7 \times 10^{-8}\,\mrm{M_{\odot}}$.
Unfortunately, also this object was too far away, and \citet{Siegert2018_V5668Sgr} determined an upper limit of $\sim 1.2 \times 10^{-8}\,\mrm{M_{\odot}}$.
A \nuc{Be}{7} ejecta mass of the order $10^{-8}\,\mrm{M_{\odot}}$ is about two orders of magnitude larger than what is theoretically expected, suggesting either a gap in the theoretical understanding or an observational bias.

Finally, the UVOIR observations of novae typically happen at the bolometric maximum luminosity which is expected to be days to weeks after the explosion \citep{GomezGomar1998}.
Short-lived isotopes such as \nuc{N}{13} ($\tau_{13} = 10\,\mrm{min}$) and \nuc{F}{18} ($\tau_{18} = 110\,\mrm{min}$), are produced during the explosive nucleosynthesis and decay by positron emission.
This suggests a strong 511\,keV line and low-energy continuum within the first few hours after the explosion -- but days to weeks before the nova had been discovered.
Therefore, only a retrospective analysis of archival data could find such a signal.
The latest of such searches has been conducted by \citet{Skinner2008_nova511_retrospective} using Swift/BAT, but resulted in no detection.
Other works that tried to discover the annihilation flash in archival data include \citet{Harris1999_novaflash,Harris2000_novaflash} using TGRS/WIND, \citet{Hernanz2000_novaflash} using CGRO/BATSE, and \citet{Smith2004_RHESSI} using RHESSI.

\subsection{Motivation and Structure}\label{sec:motivation_structure}
Searches in light curves of individual energy bands or lines limit the sensitivity of the instruments to only these particular search windows.
In addition, using extracted light curves, i.e. assuming a position in the sky but being agnostic about the temporal and spectral behavior, will bias the result and consequently the search for transients.
The most efficient use of the capabilities of $\gamma$-ray instruments is forward modeling of all available information in a complete model of the source to be analysed.
This means instead of searching for high count rates in the 511\,keV line, for example, a complete spectro-temporal model is convolved with the instrument response, and fitted together with an appropriate description of the instrumental background.
One example to implement such models is the Multi-Mission Maximum Likelihood (3ML) framework \citep{Vianello2015_3ML}, designed to perform not only multi-wavelength analyses but also multi-instrument inferences.

The other notorious problem in soft $\gamma$-ray astrophysics is the instrumental background from cosmic-ray bombardment of the spacecraft, and also the astrophysical background in all-sky monitoring instruments.
Only if these backgrounds are sufficiently well understood and modeled, weak signals, such as expected from nova explosions, can be disentangled from the raw photon count data.
In recent years, significant progress has been made in MeV astrophysics to understand the often erratic backgrounds, and utilise the instruments such as INTEGRARL/SPI \citep[cf.][]{Siegert2019_SPIBG,Siegert2021_BHMnovae} and Fermi/GBM in ways they have not been designed for.
In particular, \citet{Biltzinger2020} developed a physical background model for the the all-sky monitor GBM onboard Fermi.
This allows to study also diffuse emission and sets the basis for the analysis of hour-scale transients such as nova events.

What is missing are parametrized and public time-resolved spectra of novae that can be used with these recently developed tools to maximize the scientific return.
Therefore in this article, we will first describe our numerical schemes in the stellar evolution and Monte-Carlo radiative transfer code of $\gamma$-ray lines in Section\,\ref{sec:methods}.
In Section\,\ref{sec:evol} we present the results of our grid of nova models and discuss their dependence on model parameters.
We discuss the corresponding nucleosynthesis with focus on the radioactive isotopes in Section\,\ref{sec:nucleo}.
Section\,\ref{sec:radtran} presents our numerical results for the $\gamma$-ray light curves and spectra derived from our models.
In Section\,\ref{sec:application} we explore how the generation of the 511\,keV line, ortho-Positronium (ortho-Ps) continuum, and Compton scattered photons from novae can be utilized in a retrospective analysis of archival $\gamma$-ray data, and how our findings impact previous analyses.
Finally in Section\,\ref{sec:discuss} we compare our nucleosynthesis and radiative transfer results with representative works in the literature, discuss numerical issues and caveats of this work, and present our conclusion. 

\begin{table} 
    \centering
    \caption{The nuclear network used for nucleosynthesis of this work. The column \textit{extra} corresponds to additional isotopes outside the range provided.}
    \begin{tabular}{c c c c c}
    \hline
        element & Z & $A_{\rm min}$ & $A_{\rm max}$ & extra \\ \hline
        hydrogen & 1 & 1 & 2 & \\
        helium & 2 & 3 & 4 & \\
        lithium & 3 & 7 & 7 & \\
        beryllium & 4 & 9 & 10 & 7\\
        boron & 5 & 8 & 8 & \\
        carbon & 6 & 12 & 13 & \\
        nitrogen & 7 & 13 & 15 & \\
        oxygen & 8 & 14 & 18 & \\
        fluorine & 9 & 17 & 19 & \\
        neon & 10 & 18 & 22 & \\
        sodium & 11 & 21 & 24 & \\
        magnesium & 12 & 23 & 26 & \\
        aluminium & 13 & 25 & 27 & \\
        silicon & 14 & 27 & 28 & \\
        phosphorous & 15 & 30 & 31 & \\
        sulfur & 16 & 31 & 32 & \\ \hline
    \end{tabular}
    
    \label{table:xiso}
\end{table}

\section{Methods}\label{sec:methods}
\subsection{Stellar Evolution and Nova Simulation}\label{sec:stellar_evolution}
We use the stellar evolution code MESA version 8118 \citep{Paxton2011, Paxton2013, Paxton2015, Paxton2018, Paxton2019}.
The code solves the structure, nuclear reactions and radiative transfer inside a star with spherical symmetry.
Similar to our previous works \citep{Leung2020cow,Leung2021SN2018gep,Leung2021Wave2}, we carry out the stellar evolution and the radiative transfer calculation into two separate steps. 
The calculation is developed based on the \textit{make\_co\_wd} and \textit{nova} test cases taken from the code.
We modify the configuration files and add subroutines for this work.
%

We first use the package \textit{make\_co\_wd} to construct a CO white dwarf with a mass from 0.8--1.\,$M_{\odot}$.
The code first constructs a 3--7\,$M_{\odot}$ 
star until it develops the CO core, and then increases the optical depth of the star to enhance mass loss of the H-envelope.
Then we pass the model to the \textit{nova} package and make the white dwarf accrete H-rich matter.
We assume the matter originates from a single degenerate companion star.
The matter is mixed with $^{12}$C and $^{16}$O as impurities.
The fraction of CO-rich matter is taken as a model parameter $f_{\rm CO}$ from 10 to 50\%.
This mimics different levels of dredge-up of the CO-rich core into the H-rich envelope.
Once the H-burning runaway takes place, the code allows excited matter in super-Eddington luminosity to be ejected as wind.
We keep track of the synthesized radioactive isotopes of interest ($^{7}$Be, $^{13}$N, $^{14}$O, $^{15}$O, $^{18}$F, $^{22}$Na and $^{26}$Al) and the ejecta kinematics.
In Table\,\ref{table:xiso} we list the isotopes included for the nucleosynthesis calculations.
We limit our nuclear network here because we want to focus on the radioactive isotope production described.
The chosen network is large enough to include potential nuclear reactions for their formation while maintaining a feasible computational time.
In general, the nuclear reaction may proceed to low mass iron-group elements such as Ca \citep{Jose1998, Jose2001}.

\begin{table} 
    \centering
    \caption{Essential isotopes and their radioactive decay channels considered in this work. $Q$ is the $Q$-value of the decay.
    }
    \begin{tabular}{c|c|c|c|c}
        \hline
        isotope & half-life & $Q$ (keV) & channel & $\gamma$-rays (keV) \\ \hline
        $^{7}$Be & 53.12\,d & 477.6 & EC & $478$ \\
        $^{13}$N & 9.97\,min & 1200 & $\beta^+$ & $\leq 511$ \\
        $^{15}$O & 2.04\,min & 1735 & $\beta^+$ & $\leq 511$ \\
        $^{18}$F & 109.7\,min & 633.5 & $\beta^+$ & $\leq 511$ \\
        $^{22}$Na & 2.6\,yr & 1275 & $\beta^+$ & $\leq 511$; $1275$\\
        $^{26}$Al & 7.15\,Myr & 1809 & $\beta^+$ & $\leq 511$; $1809$ \\ \hline
    \end{tabular}
    
    \label{tab:isotopes}
\end{table}

\subsection{Monte-Carlo Radiative Transfer}\label{sec:MC_radiative}
We use the recorded data and pass it to our Monte-Carlo radiative transfer code.
The code solves the propagation, scattering and interaction of photon packets produced by the decay of radioactive nuclei.
At each time interval, we estimate the optical depth of the matter for $\gamma$-rays using a grey opacity $\kappa_{\gamma} = 0.06 Y_e$ \citep{Swartz1995}.
We choose the mass shells with an optical depth $\tau = \int_r^R \rho(r') \kappa_{\gamma} dr' = 5$ .
%
This corresponds to about $\sim$99\% reduction of the luminosity from the innermost layer considered when it arrives the surface. 
%

To generate the photon packet, we calculate the radioactive power per mass $q_{\rm decay}$ in each mass shell by including the decay of all related isotopes.
We locate the mass shells $\Delta m$ which are ejected and integrate these shells by $L_{\rm total} = \sum q_{\rm decay} \Delta m$ to find the instantaneous $\gamma$-ray emission. 
They correspond to a collection of photons undergoing the same process during their propagation inside the star \citep{Ambwani1988}.
Instead of splitting into multiple packets when scattering occurs, they can lose energy.
Each packet therefore corresponds to the `instantaneous photons emitted per unit time' instead of a real photon.
Thus, a packet carries a luminosity $L_{\rm packet}$ that corresponds to $L_{\rm total} / N_{\rm packet}$.
In general we find that the time delay from photon emission to escape is short ($\sim 1$\,ms or lower) so that the static approximation is appropriate.

\subsubsection{Photon Generation}\label{sec:photon_generation}
The code assumes that the $\gamma$-rays originate solely from the decay of radioactive isotopes and we calculate the spectra from given time snapshots of the nova model during its mass ejection.
We put particular emphasis on the first day of the mass ejection as we are interested in the decays of $^{13}$N and $^{18}$F, which have short half life times compared with the expansion time scale of the nova ($\sim 10$ days for our models).
These isotopes are distinctive from the others because their main decay channel is $\beta^+$-decay, which emits a positron or in addition to (in the case of \nuc{O}{15}) an energetic photon.
In Table\,\ref{tab:isotopes} we tabulate the principle parameters for the isotopes of interest.

If an isotope decays and subsequently emits a photon, a photon packet is created by assigning it the corresponding $\gamma$-ray line energy and an arbitrary direction.
The $\gamma$-ray energies are taken from the co-moving frame with the moving nuclei.
In our calculation, the thermal fluctuation and collective motion of the nuclei is small enough that the Doppler effect is negligible. 

When an isotope undergoes $\beta^+$-decay, the positron quickly loses its energy through Coulomb interaction and captures an electron to form Positronium (Ps). 
%
%
%
Ps has two spin states: para-Ps and ortho-Ps.
The former emits two photons and the latter three due to charge and spin conservation \citep[see][for fundamental features of Ps]{Ore1949_511,Berko1980}.
Quantum statistics limits their ratio to a maximum of para:ortho$=$1:4.5.
The exact ratio depends strongly on the matter density and temperature \citep[e.g.,][]{Leising1987}.
In our case, where the matter is opaque and dense, we expect the ratio to approach the quantum limit.
When two-photon emission occurs, we assign two photon packets of energy 511\,keV.
The first one has an arbitrary direction, with the direction of the second packet chosen by momentum conservation.
When three-photon emission occurs, the individual photon energy is set by Monte-Carlo process which reproduce the observed probability distribution, and the sum of the three photons conserves the total energy 1022\,keV and the initial momentum.
%
Then we assign an arbitrary direction for one photon packet and obtain the directions for the other two photon packets by conservation of momentum.
The time-delay from the formation of Ps to its decay is $10^{-9}$--$10^{-6}$\,s \citep{Czarnecki1999}, which is much shorter than its escape time.

\begin{figure*}
    \centering
    \includegraphics[width=0.49\textwidth,trim=0.3in 0.0in 0.5in 0.2in, clip=true]{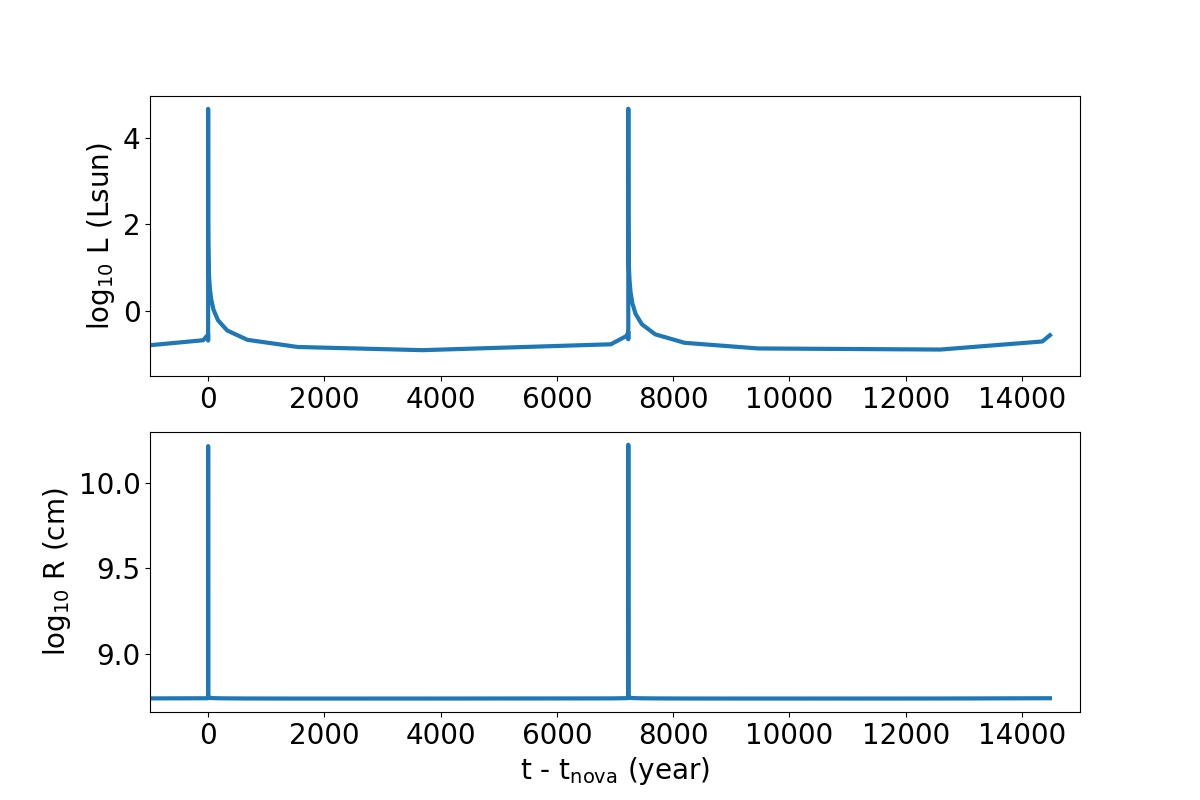}
    \includegraphics[width=0.49\textwidth,trim=0.3in 0.0in 0.5in 0.2in, clip=true]{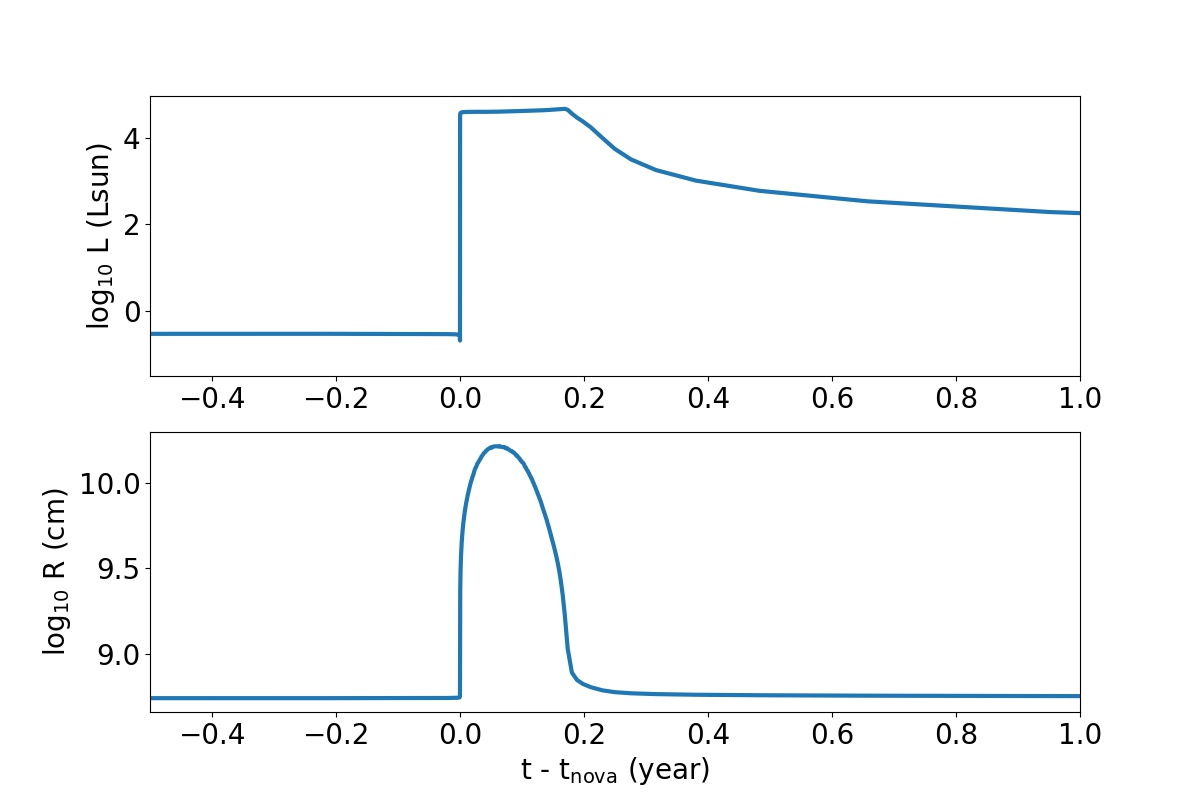}
    \caption{(left) Luminosity and radius evolution for the nova model during its first two pulses with $M = 1.0\,M_{\odot}$ and CO fraction of 50\%. (right) Zoomed into one year around the first pulse. 
    }
    \label{fig:L_R_plot}
\end{figure*}

\subsubsection{Photon Interactions}\label{sec:photon_interactions}
We consider the following three types of interaction processes that change the energy of a photon packet \citep{Pozdnyakov1983}: 

\noindent (1) Compton scattering: the photon transfers energy to an electron and loses energy.
The energy before $E$ and after $E'$ the interaction is related by:
\begin{equation}
    E' = \frac{E}{1 + (E/m_e c^2) \cos \theta}\mrm{,}
    \label{eq:compton_scattering}
\end{equation}
where $m_e$ is the electron mass and $\theta$ is the scattered angle in the center-of-mass frame. \newline 

\noindent (2) Photopair production: the photon packet is assumed to lose all its energy to an electron and the electron later emits an $e^-$-$e^+$-pair similar to the two-photon emission described above.
The cross section of this process is calculated according to the photon energy:
    \begin{eqnarray}
        \sigma &=& A (E - 1.022) Z^2 \times 10^{-27} {\rm cm^2}, ~ 1.022 < E < 1.5 {\rm MeV}\mrm{;} \nonumber \\ 
        \sigma &=& [B_1 + B_2 (E - 1.5)] Z^2 \times 10^{-27} {\rm cm^2}, ~ E \geq 1.5 {\rm MeV}\mrm{,}
        \label{eq:photopair_production}
    \end{eqnarray}
with $A$, $B_1$ and $B_2$ being 1.0063, 0.481 and 0.301 respectively.
$E$ is the photon packet energy in units of MeV. \newline

\noindent (3) Photoelectric absorption: the photon is assumed to be absorbed by a nucleus.
The cross section takes the equation:
\begin{equation}
    \sigma = C (E / {\rm 100~keV})^{-3.2} \times 10^{-24} {\rm cm^2}\mrm{,}
    \label{eq:photoeffect}
\end{equation}
with $C = 0.0448$ for elements with $Z \sim 7$. 
Although the exact value of $C$ depends on the atomic number, $\sigma \sim E^{-3}$ holds true for a wide range of elements, which means photoelectric absorption dominates the destruction of photon packets as the photon energy reaches below $\sim 100$\,keV.


We assume that the photon has at least one interaction when it travels across $\delta \tau = 1$.
To determine which process takes place, in each step, we assign a random number $\Delta \tau \in (0,1)$ which corresponds to the distance traveled by the photon packet before the next interaction.
The mean free path $\Delta r_i = \Delta \tau / (\rho \kappa_i)$ for $\kappa_i$ is taken from the three processes described above.
By finding the minimum $\Delta r_i$, we assign the corresponding probability for each process to take place.
Then we generate another random number to decide which process is chosen.
In the case where the traveling distance crosses the mass shell (defined by the stellar evolution model), we also update the local thermodynamical properties experienced by the photon packet. 

\section{Evolution of Novae}\label{sec:evol}
We generate a grid of nova models by varying the white dwarf (WD) mass $M$ (0.8--1.2\,$M_{\odot}$), CO fraction $f_{\rm CO}$ (10--50\%), mass accretion rate $\dot{M}$ and convection parameters. 
In Figure\,\ref{fig:L_R_plot} we show the luminosity and radius as a function of time for $M = 1.0\,M_{\odot}$, $f_{\rm CO} = 0.5$ and $\dot{M} = 10^{-9}\,M_{\odot}$ yr$^{-1}$.
We plot two pulses in the evolution of a recurrent nova.
During the nova outburst, the luminosity can increase by five orders of magnitude and reach $\sim 10^{4}\,L_{\odot}$.
The WD radius also increases by one order of magnitude during the nova outburst.
The two pulses are separated by about 7000 years.
The exact timing depends on the mass accretion rate, the mass of the WD and mixing of accreted matter.
For recurrent novae, the mass accumulated before the TNR occurs is roughly similar.
In the right panel of the same plot, we magnify the first nova outburst.
Zooming into the first outburst, it becomes clear that the whole outburst lasts about 0.2\,yr until most thermally excited matter is ejected.
It takes a much longer time ($\sim 10$--$100$ years) for the nova to relax and to restore its original luminosity.
The decline of the luminosity satisfies $L = L_0 t^{-1}$, suggesting the WD cools by blackbody radiation. 
The decrease of the radius happens because the super-Eddington wind gradually removes the hot matter in the outer parts of the nova.
The mass ejection is efficient in removing the deposited matter until it reaches its original mass before mass accretion. 

\begin{figure*}
    \centering
    \includegraphics[width=16cm,trim=1.0in 0.5in 1.0in 1.0in,clip=True]{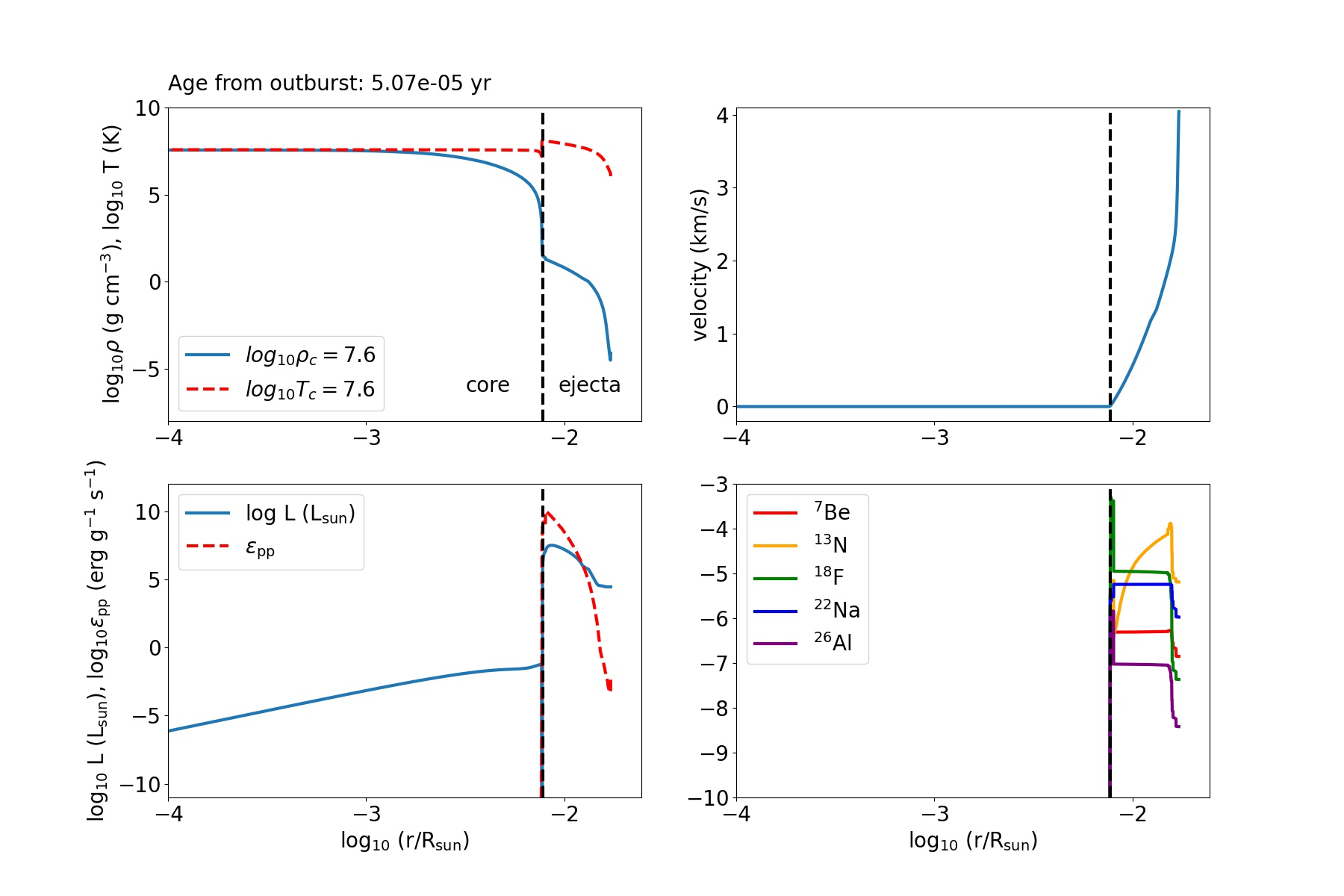}    \caption{(top left) Density and temperature profile for the nova model with $M = 1.0\,M_{\odot}$, $f_{\rm CO} = 0.5$, $\dot{M} = 10^{-9}\,M_{\odot}$\,yr$^{-1}$ before the super-Eddington wind at $\sim 10^{-5}$ yr after the TNR (first in peak in Fig.\,\ref{fig:L_R_plot})
    as a function of white dwarf radius. (top right) Velocity profile. (bottom left) Luminosity and \textit{pp}-chain energy production rate. (bottom right) Isotope abundances for selected radioactive isotopes. 
    }
    \label{fig:star_profil}
\end{figure*}

In Figure\,\ref{fig:star_profil} we show the stellar profiles for the same nova model as above, focusing on the thermodynamics and kinematics.
The snapshot is taken before the wind mass loss removes the thermally excited matter.

The nova consists of two layers, the compact CO core with a radius $\sim 10^{-2}\,R_{\odot}$ and the sparse accreted H-rich matter which extends to $\sim 10^{-1}\,R_{\odot}$.
The temperature bump near $r = 10^{-2}\,R_{\odot}$ also shows the location where explosive H-burning has happened.
This is consistent with the luminosity profile, also being increased at the interface and gradually decreasing as the nuclear reactions occur in a mass shell with a small mass $\sim 10^{-5}\,M_{\odot}$. 
From the velocity profile, only the outer part of the H-envelope is expanding, but with a very low velocity $\sim 1$--$3\,\mrm{km\,s^{-1}}$.
The expansion does not lead to a dynamical mass outburst because the typical escape velocity is $\sim 10^3\,\mrm{km\,s^{-1}}$.
Instead, the nova loses its matter by radiative winds.
During TNR of the H-envelope, the heat creates strong convection, which allows the produced radioactive isotopes to be efficiently transported and some $\gamma$-rays to escape easily.
Only those near the surface are important for the early $\gamma$-ray signal because of the very short lifetime of $^{13}$N and $^{18}$F.

\begin{figure}
    \centering
    \includegraphics[width=8cm,trim=0.0in 0.0in 1.0in 0.5in,clip=True]{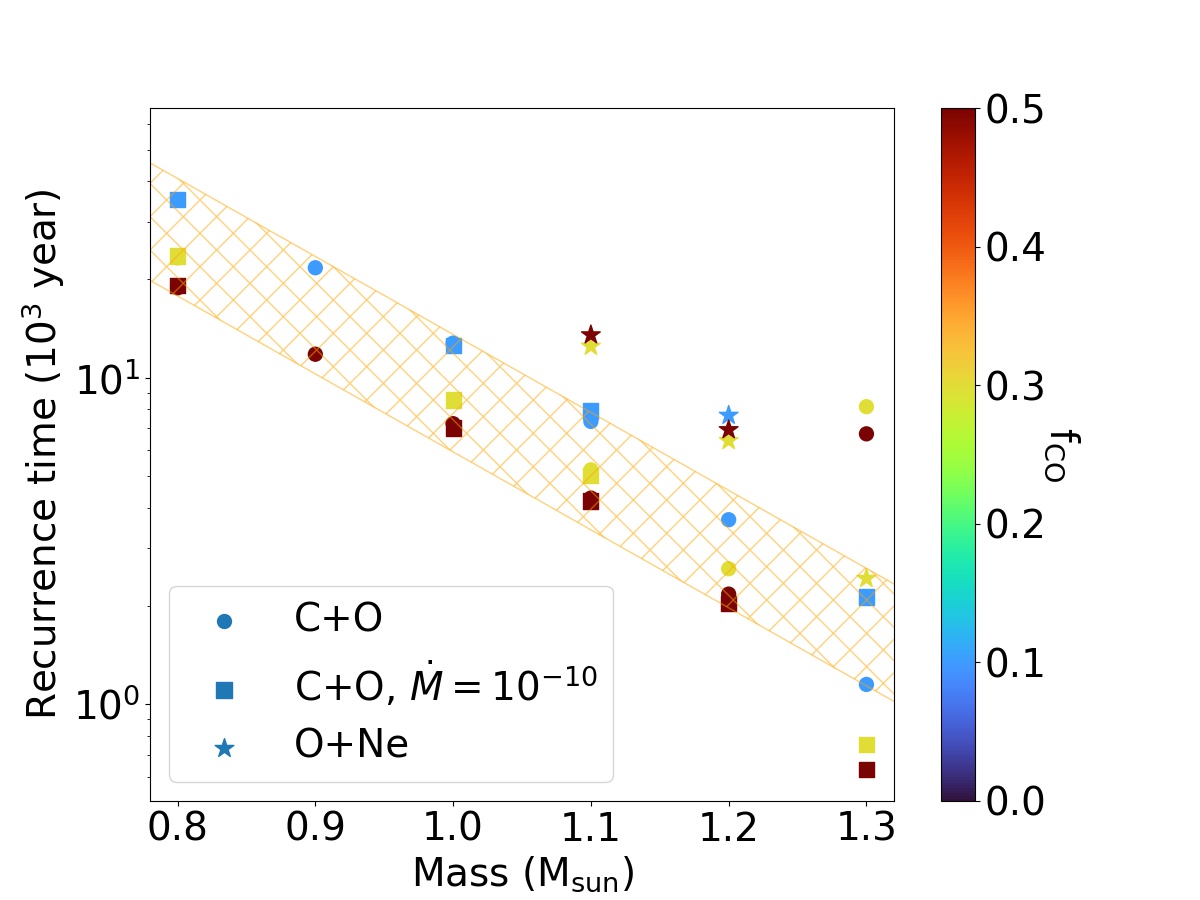}
    \caption{Recurrence time of nova models assuming a CO WD with $\dot{M} = 10^{-9}\,M_{\odot}$\,yr$^{-1}$ (circles), a CO WD with $\dot{M} = 10^{-10}\,M_{\odot}$\,yr$^{-1}$ (squares) and an ONe WD with $\dot{M} = 10^{-9}\,M_{\odot}$\,yr$^{-1}$ (stars). Colors represent the mixing ratio. The recurrence time for the models with $\dot{M} = 10^{-10}\,M_{\odot}$\,yr$^{-1}$ are scaled down by a factor of 10. The orange hatched area corresponds to the empirical fitting function from Table \ref{table:fit}.
    }
    \label{fig:recurr_time}
\end{figure}

We surveyed the nova models by spanning the progenitor WD mass and the mixing efficiency.
In Figure\,\ref{fig:recurr_time} we show the recurrence times for novae as depending on the mixing, accretion rate, and WD composition.
The recurrence time is defined by the time lapse between the first and second outburst.
The models show a tight exponential relation for models with 0.8--1.2\,$M_{\odot}$ for CO WDs.
Beyond 1.2 $M_{\odot}$, the degeneracy leads to models deviating from this trend.
A similar relation is observed for ONe WDs between 1.1--1.2\,$M_{\odot}$.

\begin{figure*}
    \centering
    \includegraphics[width=8cm,trim=0.1in 0.1in 0.5in 0.5in,clip=True]{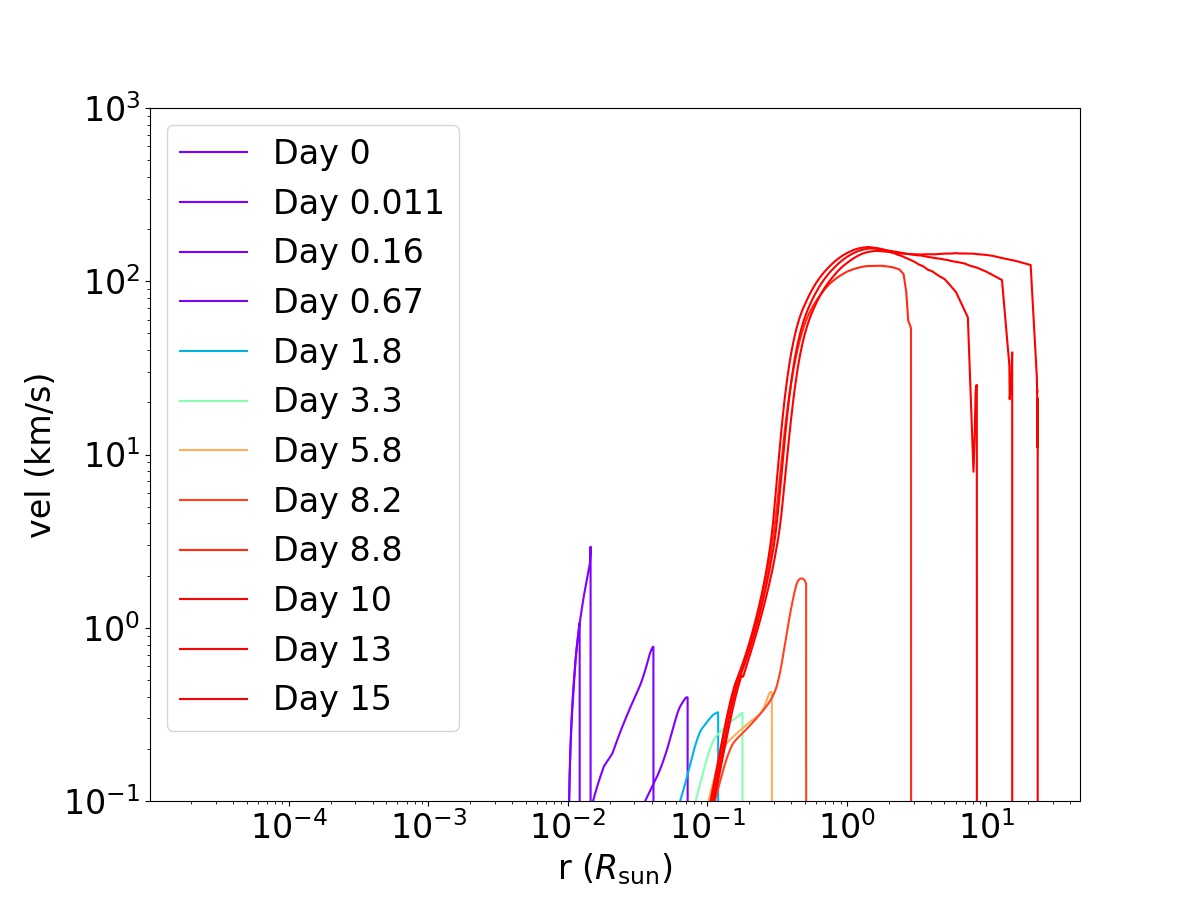}
    \includegraphics[width=8cm,trim=0.1in 0.1in 0.5in 0.5in,clip=True]{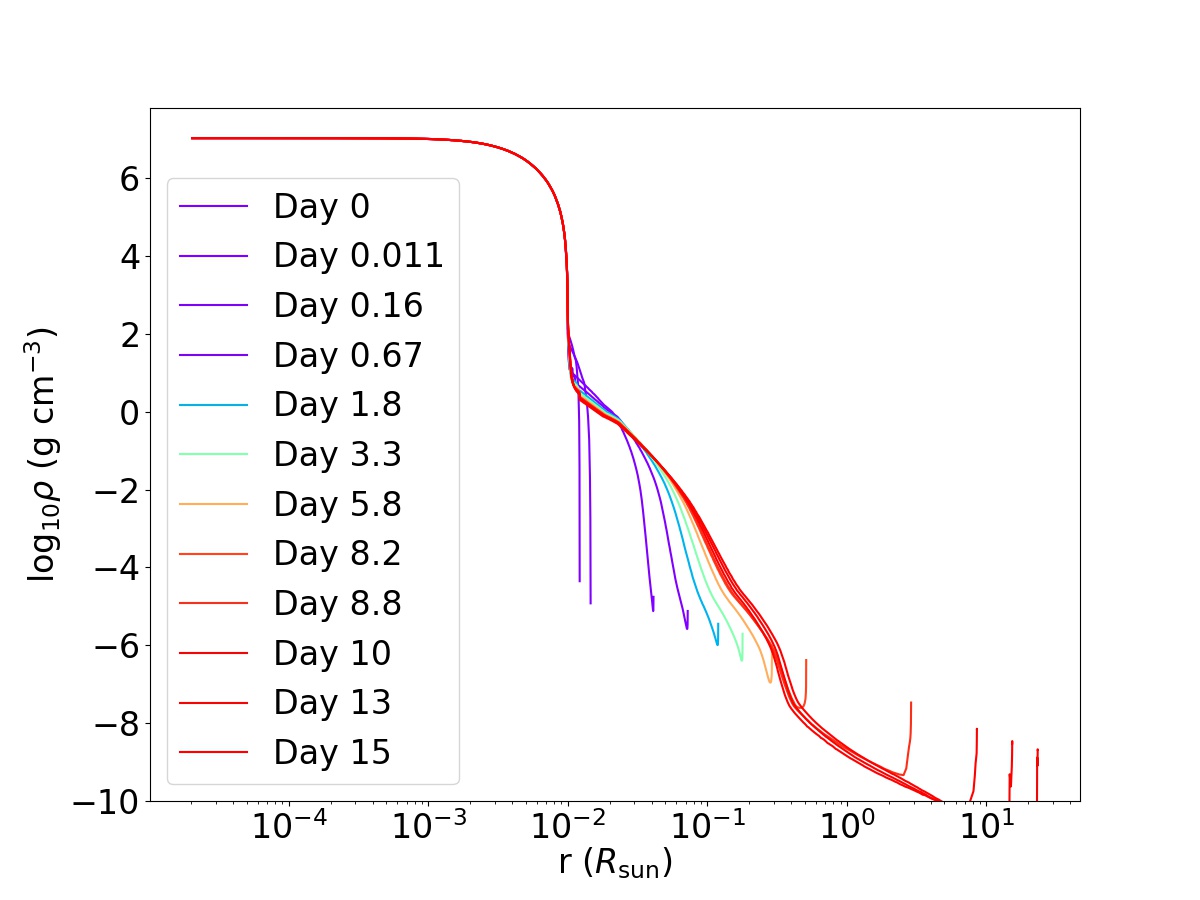}
    \includegraphics[width=8cm,trim=0.1in 0.1in 0.5in 0.5in,clip=True]{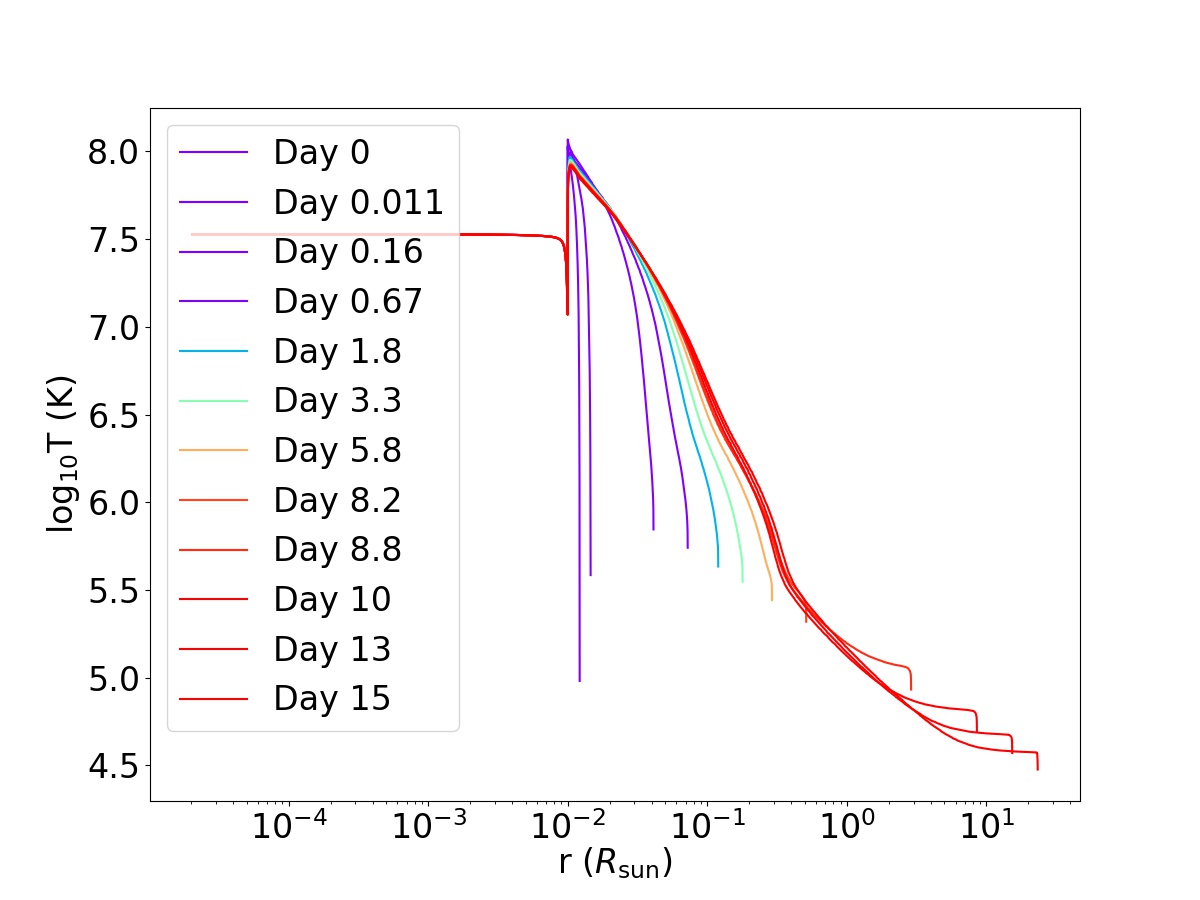}
    \includegraphics[width=8cm,trim=0.1in 0.1in 0.5in 0.5in,clip=True]{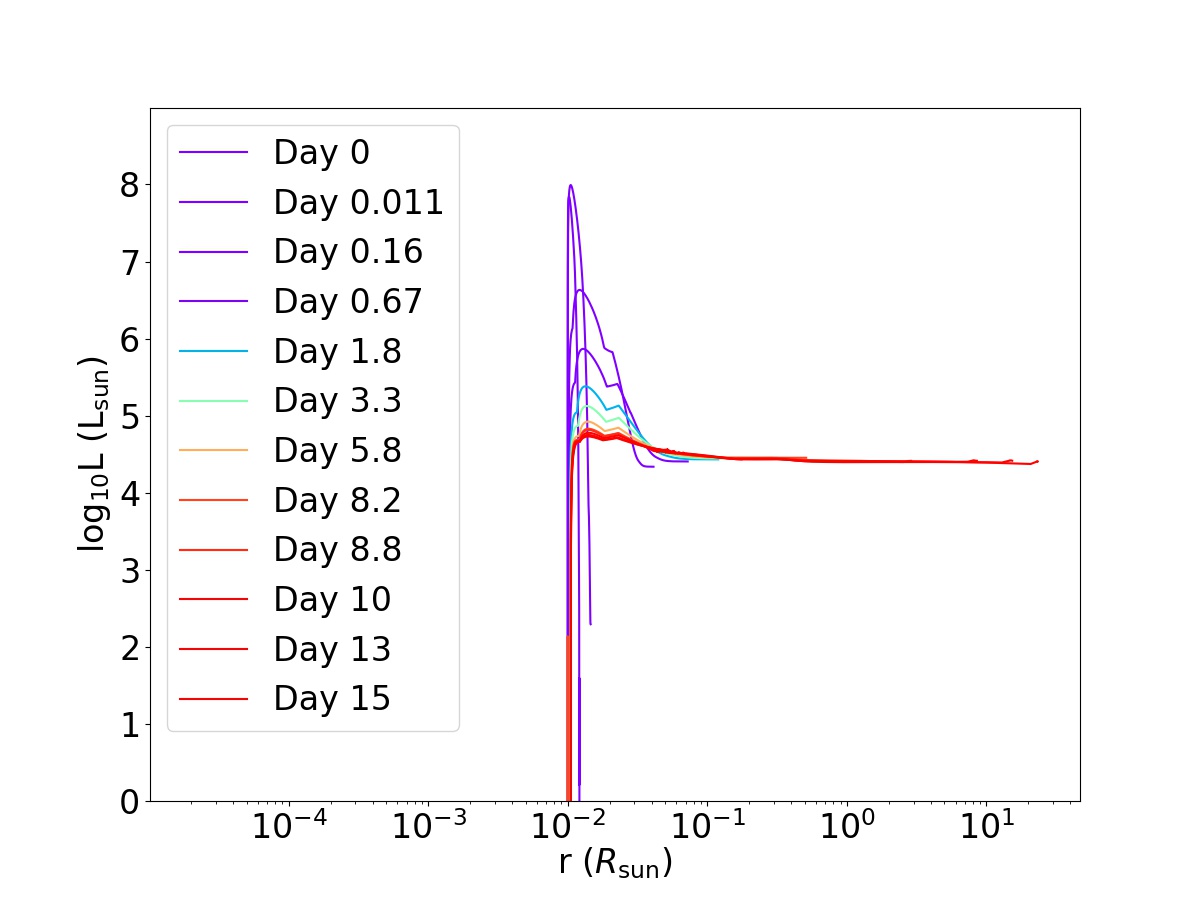}
    \caption{(top left) Velocity profiles of nova ejecta at selected times from the CO WD model $M = 0.80\,M_{\odot}$, with $\dot{M} = 10^{-9}\,M_{\odot}$\,yr$^{-1}$. (top right) Density profiles. (bottom left) Temperature profiles. (bottom right) Luminosity profiles. 
    }
    \label{fig:hydro_M080}
\end{figure*}

We determine a smooth relation of the recurrence times by fitting the function $t_{\rm recur} = A \exp(-B M)$, where $M$ is the WD mass, and $A$ and $B$ are fitted constants as listed in Table\,\ref{table:fit}.
The fitting suggests that $B \sim -5.5$ for a wide range of $f_{\rm CO}$.
%
%
The choice of the mass accretion rate does not change the trend of the models.
An exception occurs for the CO WD model with $\dot{M} = 10^{-10}\,M_{\odot}$\,yr$^{-1}$ and $M = 1.3\,M_{\odot}$, which has an exceptionally high recurrence time.
The ONe WDs also show a similar trend which can be well described by the same scaling.

\begin{figure}
    \centering
    \includegraphics[width=8cm,trim=0.1in 0.1in 0.5in 0.5in,clip=True]{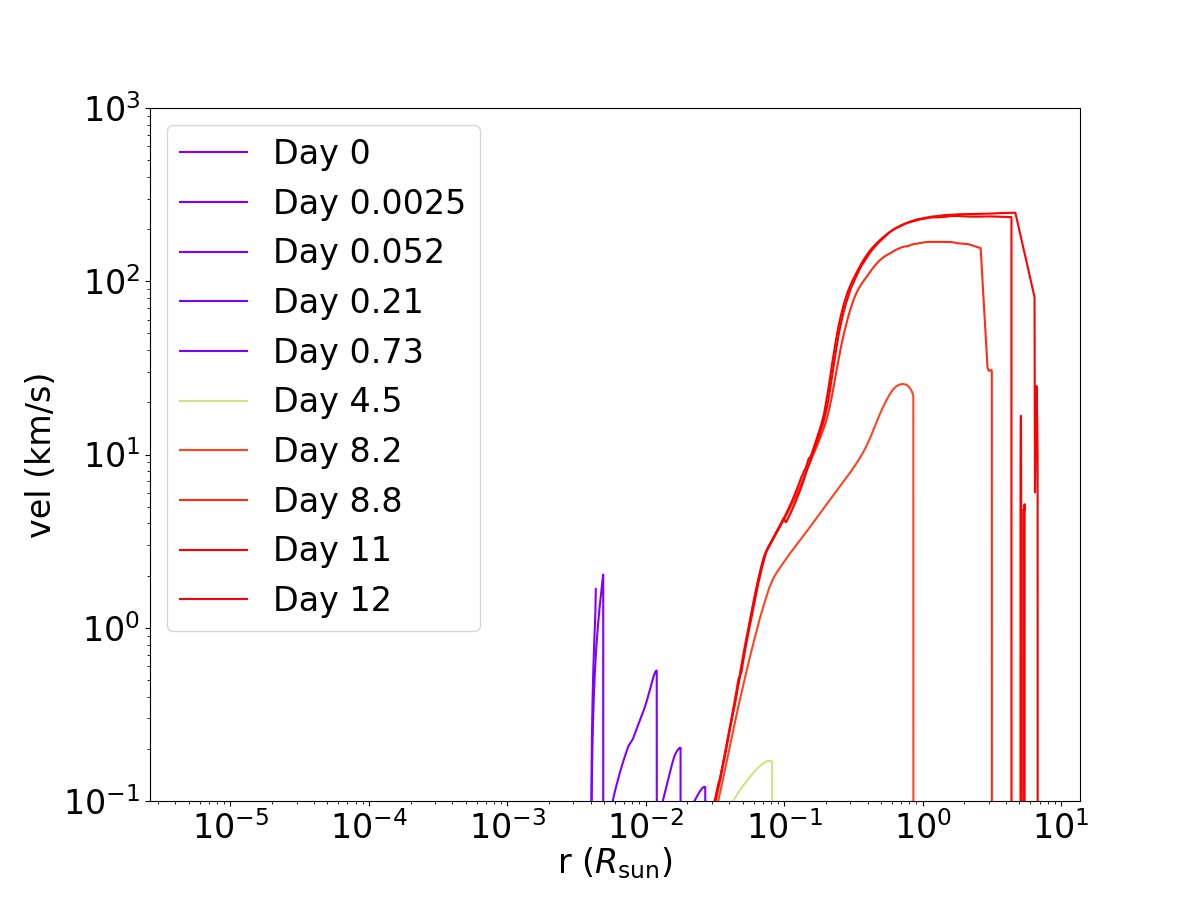}
    \caption{Same as Figure\,\ref{fig:hydro_M080} but for the model with $M = 1.30\,M_{\odot}$. 
    }
    \label{fig:hydro_M130}
\end{figure}

To estimate the expansion velocity, we show the velocity and density profiles as a function of time.
For this part, we repeat the model but suppressed all super-Eddington mass loss so that we can follow how the ejecta develop toward their asymptotic velocity during expansion.
As an example, we show the nova models of a CO WD with $\dot{M} = 10^{-9}\,M_{\odot}$\,yr$^{-1}$, $f_{\rm CO} = 0.5$, and a mass of $M = 0.8\,M_{\odot}$ and $M = 1.3\,M_{\odot}$, respectively, in Figures\,\ref{fig:hydro_M080} and \ref{fig:hydro_M130}.

In the first model it takes about 8--10 days for the motion to reach the surface and develop an expansion flow similar to homologous expansion.
At early times ($\lesssim 8$\,d), the ejecta have a low velocity of $\sim 1\,\mrm{km\,s^{-1}}$ and are decelerating with time, showing that they remain bounded by gravity.
Only around Day 8 when the momentum reaches the low density tail of the ejecta, they gradually develop the rapid expansion with an asymptotic velocity $200$--$300\,\mrm{km\,s^{-1}}$.
Once such a velocity is reached, the ejecta no longer accelerate.
Higher mass models have a stronger deceleration at early times, but the density structure for the low density tail is similar.
Both profiles suggest that the mass ejection process is similar regardless of the progenitor.
The slow velocity at early times also suggests that the ejecta with radioactive isotopes remain compact for a time longer than the half lives of, for example, $^{13}$N and $^{18}$F.
Such an opaque environment largely suppresses the propagation of $\gamma$-rays.

In the temperature and luminosity profiles, we observe that during the TNR of H, the peak temperature can reach as high as $10^{8.2}$\,K.
This provides the necessary thermal pressure for the expansion, where we see the long tail of matter with its temperature falling to about $\sim 10^{5}$ K\,around Day 10.
The high temperature allows most elements in the ejecta to remain ionized.
The peak of nuclear reactions can raise the luminosity to about $10^8\,L_{\odot}$.
After that the luminosity quickly falls to a constant level of about $10^{4.5}\,L_{\odot}$ when the surface matter expands. 

\begin{table}
    \centering
    \caption{Estimated parameters for the recurrence time of nova models as a function of mixing and accretion rate. 
    }
    \begin{tabular}{c|c|c|c}
        \hline
         $f_{\rm CO}$ &  $\dot{M}$ ($M_{\odot}$ yr$^{-1}$) & $A$ & $B$ \\ \hline
         
         0.1 & $10^{-9}$ & $3.32 \times 10^{6}$ & -5.5 \\
         0.3 & $10^{-9}$ & $1.85 \times 10^{6}$ & -5.5 \\
         0.5 & $10^{-9}$ & $1.45 \times 10^{6}$ & -5.5 \\ \hline
    \end{tabular}
    \label{table:fit}
\end{table}

To understand how the matter ejection depends on the WD progenitor, we examine the velocity evolution of the near-Chandrasekhar mass CO WD model with $M = 1.3\,M_{\odot}$.
The evolution is almost identical to the lower mass models despite the large difference in mass and initial radius.
The asymptotic velocity is slightly higher, about $300\,\mrm{km\,s^{-1}}$, but is marginally below the escape velocity.
The ejecta maintain a high velocity at Day 10 when it reaches a radius of $10\,R_{\odot}$, suggesting that they will become unbounded at later times.
Again, the development of the velocity flow takes about 9 days before the velocity reaches its asymptotic value.
Thus, the short lifetime radioactive isotopes, such as \nuc{N}{13} and \nuc{F}{18} -- the drivers for the annihilation flash -- have decayed before the matter becomes transparent to $\gamma$-rays.

To further explore the trend of the mass deposition and ejection of the nova events, we plot the total ejected mass in Figure\,\ref{fig:models_Mej}.
The models show a clear falling trend with increasing progenitor WD mass.
The typical ejected mass decreases from $\sim 2$--$3 \times 10^{-5}\,M_{\odot}$ for WDs with masses of 0.8\,$M_{\odot}$ to $\sim 5$--$10 \times 10^{-7}\,M_{\odot}$ for WDs near the Chandrasekhar mass.
The models show a small variation among different mixing efficiencies.
A similar falling trend is observed for ONe WDs with a tenfold larger ejecta mass compared to CO WDs with the same mass.

We show the global maximum temperature of the models during the outburst in Figure\,\ref{fig:models_Tmax}.
Contrasted to the ejected mass, the maximum temperature experienced in the star increases with the progenitor mass approximately exponentially.
The variation among models for the same mass is small, within $\sim 0.05$ in logarithmic scale.
The maximum temperature always occurs at the bottom of the H-burning zones.
A higher maximum temperature is expected because the corresponding density for the H-burning shells is higher for a higher mass WD.
As a result, the nuclear reactions can proceed more readily once the temperature is sufficiently high.
ONe WDs also show an increasing trend with a similar slope, however at higher temperatures around $220$\,MK.
%
%
Such a high temperature may allow more radioactive isotopes to be synthesized.

\section{Nucleosynthesis in Novae}\label{sec:nucleo}
The stellar evolution models with the extended nuclear reaction network allows us to compute explicitly the detailed chemical composition, and especially those of radioactive isotopes, in the ejected matter.
In each model, we capture the chemical abundance before the super-Eddington wind mass loss begins.
This ensures that the all synthesized metal is properly captured.

\subsection{Productions of Radioactive Isotopes}\label{sec:radioactive_isotopes}
\begin{figure*}
    \centering
    \includegraphics[width=8cm,trim=0.0in 0.0in 0.5in 0.9in,clip=True]{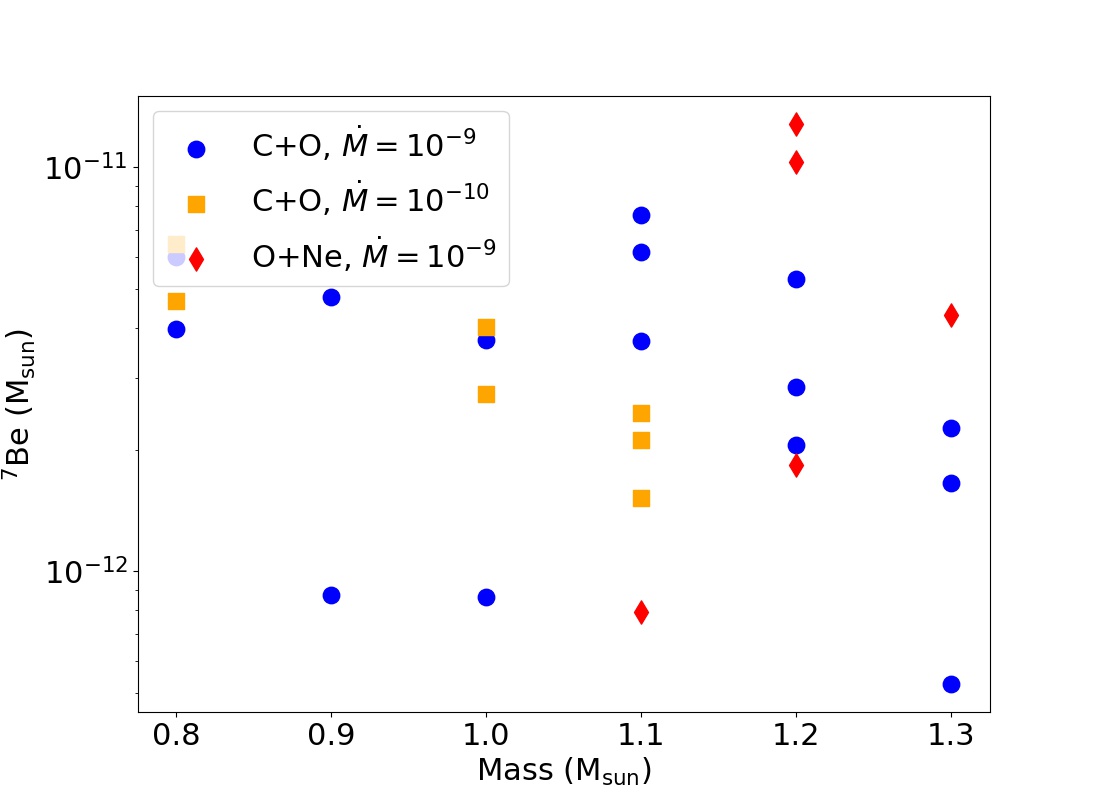}
    \includegraphics[width=8cm,trim=0.0in 0.0in 0.5in 0.9in,clip=True]{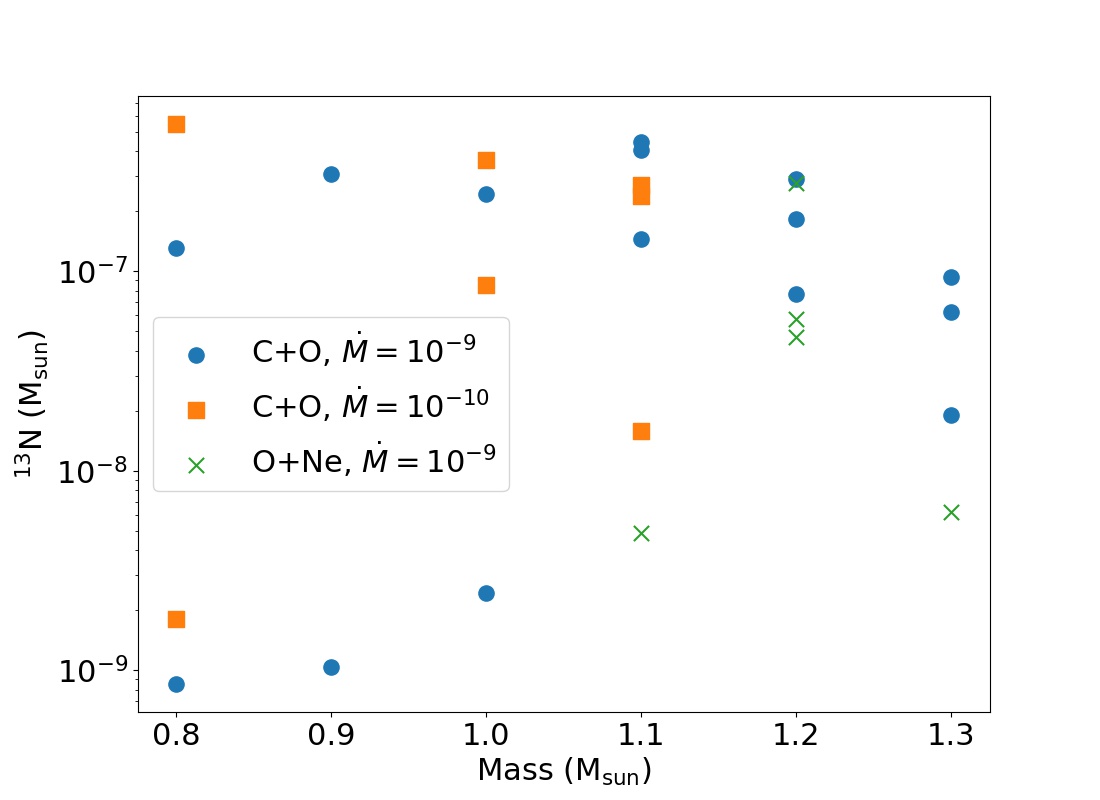}
    \includegraphics[width=8cm,trim=0.0in 0.0in 0.5in 0.9in,clip=True]{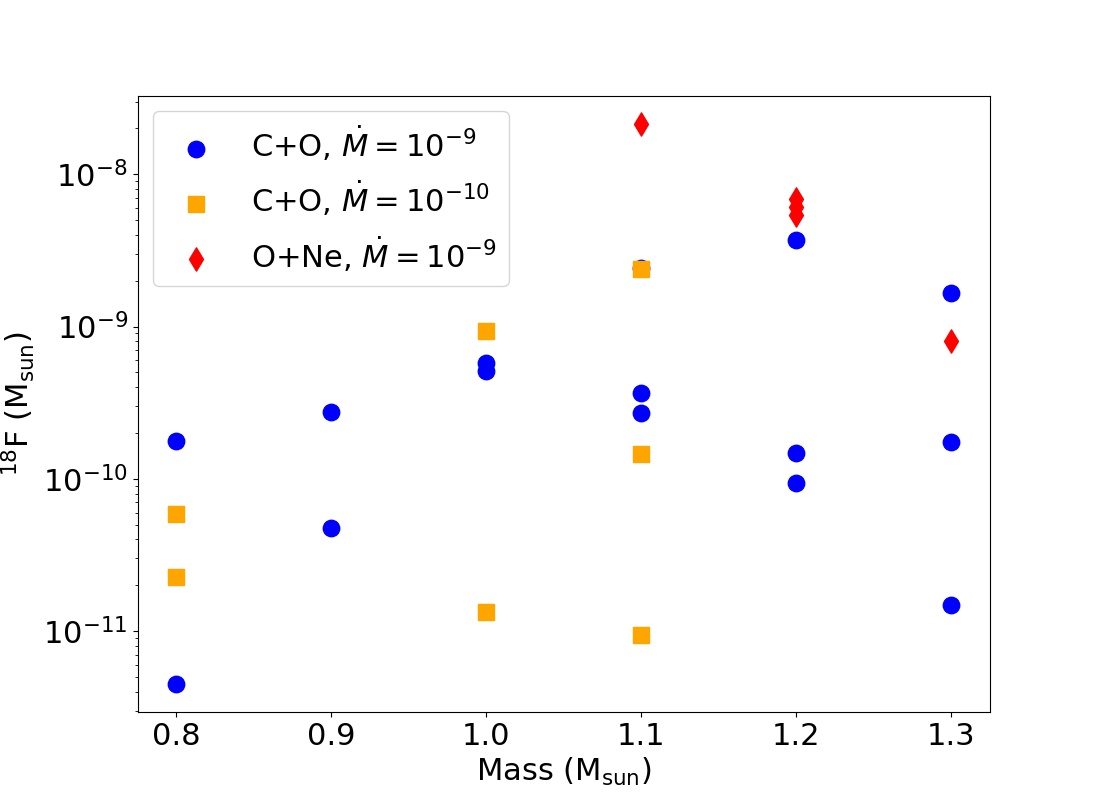}
    \includegraphics[width=8cm,trim=0.0in 0.0in 0.5in 0.9in,clip=True]{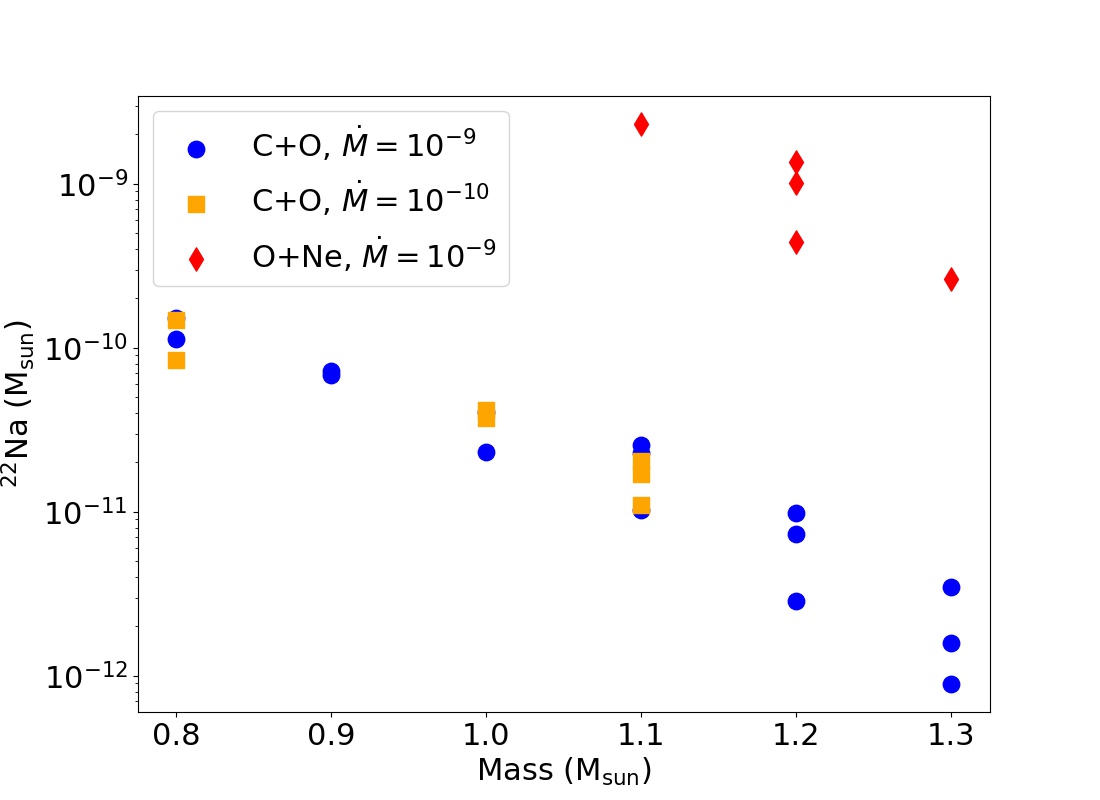}
    \caption{(top left) Total synthesized $^{7}$Be in the ejecta before its mass ejection occurs. (top right) $^{13}$N. (bottom left) $^{18}$F. (bottom right) $^{22}$Na. 
    }
    \label{fig:nova_nucleo}
\end{figure*}

In Figure\,\ref{fig:nova_nucleo} we show the nucleosynthesis results of our nova models.
%
The models are classified based on the composition (CO or ONe) and the mass accretion rate from its companion stars.

It is evident that there is no clear trend for the lighter isotopes including $^{7}$Be, $^{13}$N and $^{18}$F.
The scatter of the mass can range from one to three orders of magnitude and without a consistent mass dependence.
$^{7}$Be has a narrower range from $10^{-12}$--$10^{-11}\,M_{\odot}$, while $^{13}$N ($^{18}$F) can range between $10^{-9}$--$10^{-6}\,M_{\odot}$ ($10^{-8}$--$10^{-5}\,M_{\odot}$).
There is also no specific trend with the WD composition.
On the other hand, there is a clear decreasing trend for $^{22}$Na as a function of WD mass.
ONe WDs have a similar trend but with a typical mass about 10--100 times higher than CO WDs ($10^{-12}$--$10^{-10}\,M_{\odot}$). 

\begin{figure}
    \centering
    \includegraphics[width=8cm,trim=0.0in 0.0in 0.5in 0.9in,clip=True]{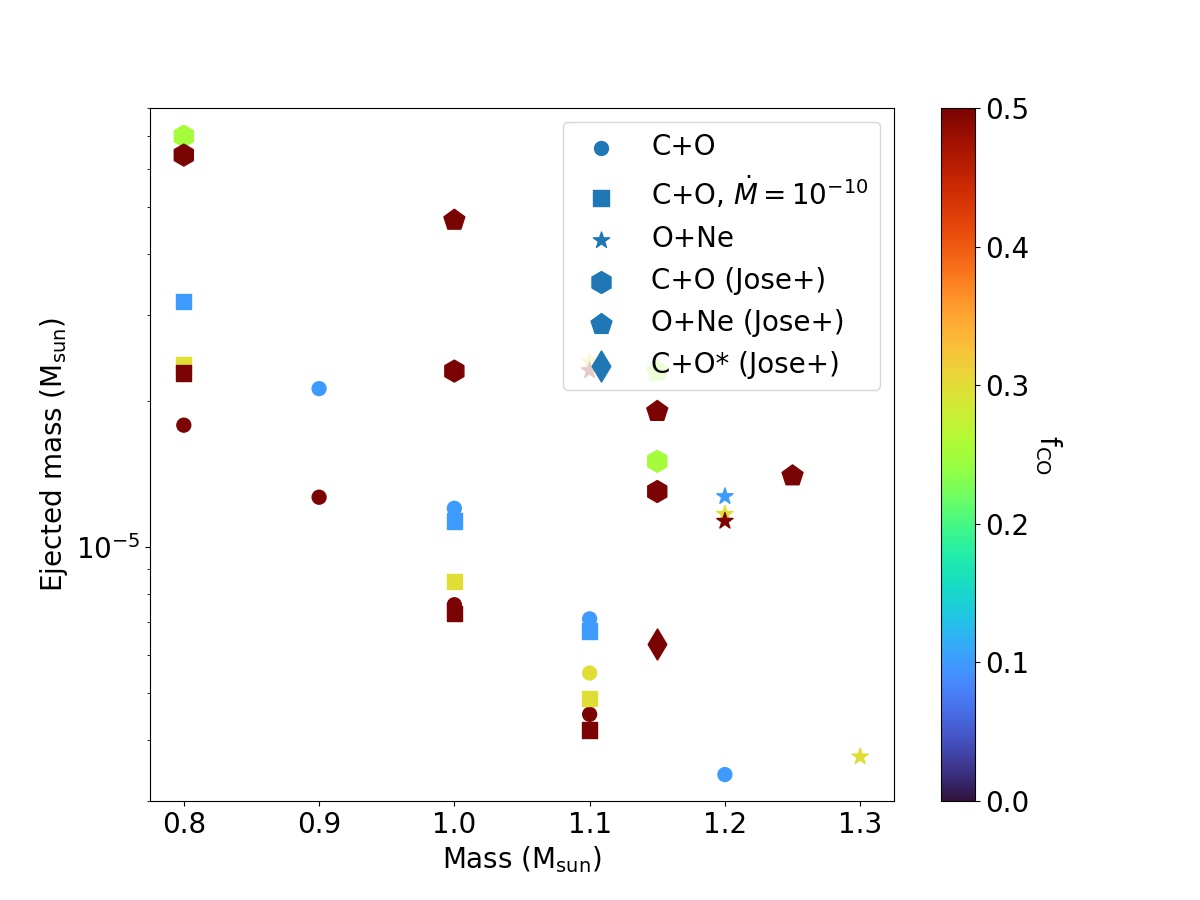}
    \caption{Total ejected mass during nova outbursts for models explored in this work, including CO WDs with $\dot{M}=10^{-9}\,M_{\odot}$\,yr$^{-1}$ (circle), $\dot{M}=10^{-10}\,M_{\odot}$\,yr$^{-1}$ (square) and ONe WDs with $\dot{M}=10^{-9}\,M_{\odot}$\,yr$^{-1}$ (star). Jose+ stands for models taken from \citet{Jose1998}.}
    \label{fig:models_Mej}
\end{figure}

\begin{figure}
    \centering
    \includegraphics[width=8cm,trim=0.0in 0.0in 0.5in 0.9in,clip=True]{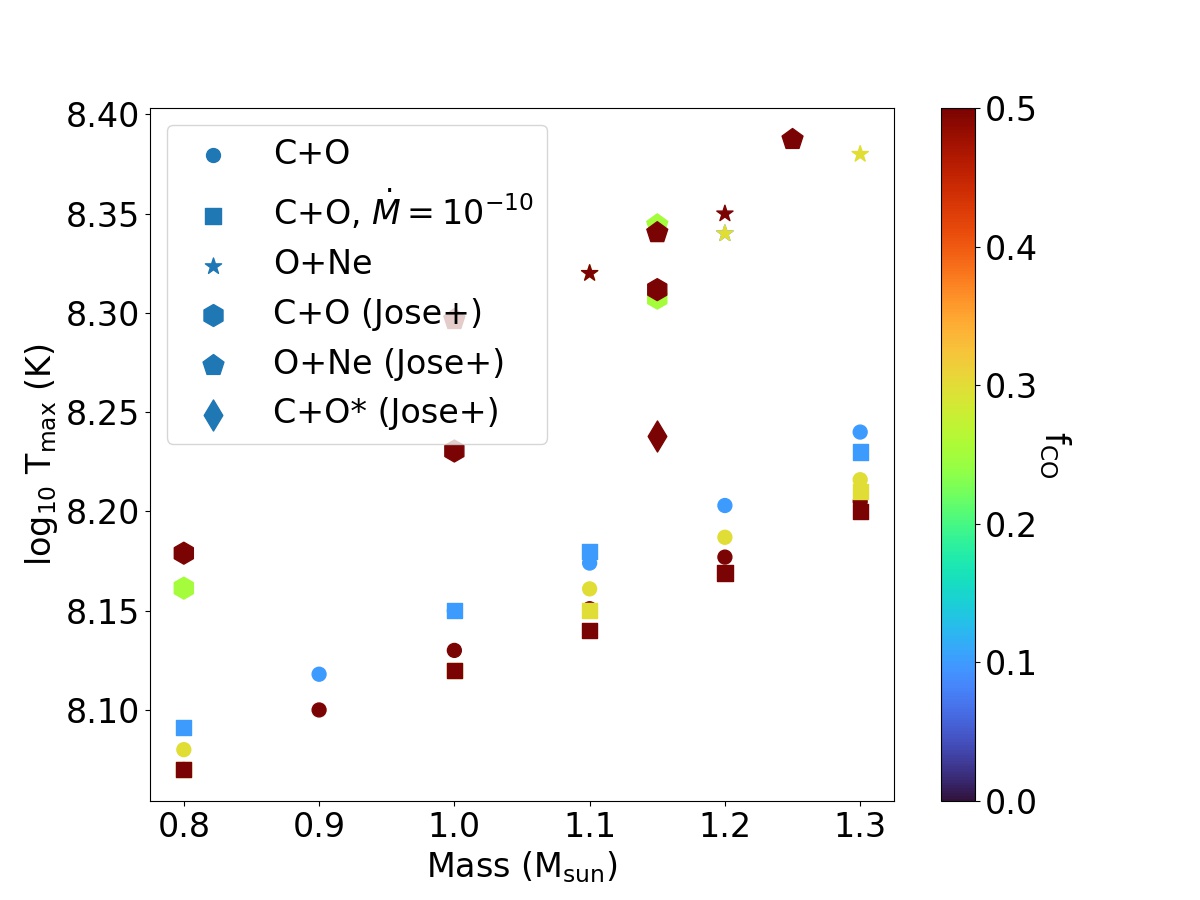}
    \caption{Maximum temperature during H-nuclear runaway for models explored in this work, including CO WDs with $\dot{M}=10^{-9}\,M_{\odot}$\,yr$^{-1}$ (circle), $\dot{M}=10^{-10}\,M_{\odot}$\,yr$^{-1}$ (square) and ONe WDs with $\dot{M}=10^{-9}\,M_{\odot}$\,yr$^{-1}$ (star). 
    }
    \label{fig:models_Tmax}
\end{figure}

\newpage
\subsection{Production of Stable Isotopes}\label{sec:stable_isotopes}

\begin{figure*}
    \centering
    \includegraphics[width=5.75cm,trim=0.0in 0.0in 0.5in 0.6in,clip=True]{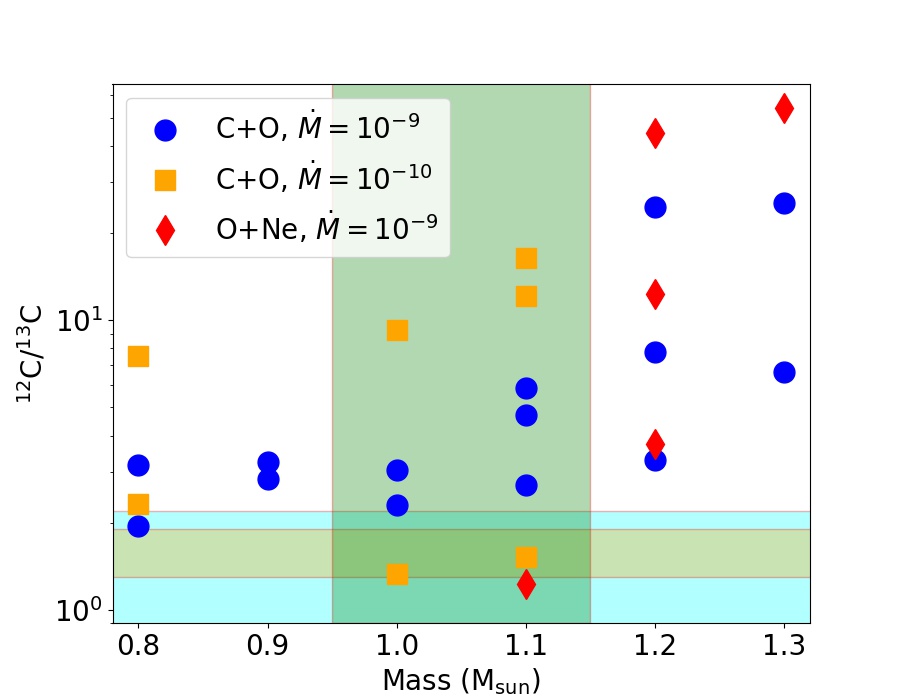}
    \includegraphics[width=5.75cm,trim=0.0in 0.0in 0.5in 0.6in,clip=True]{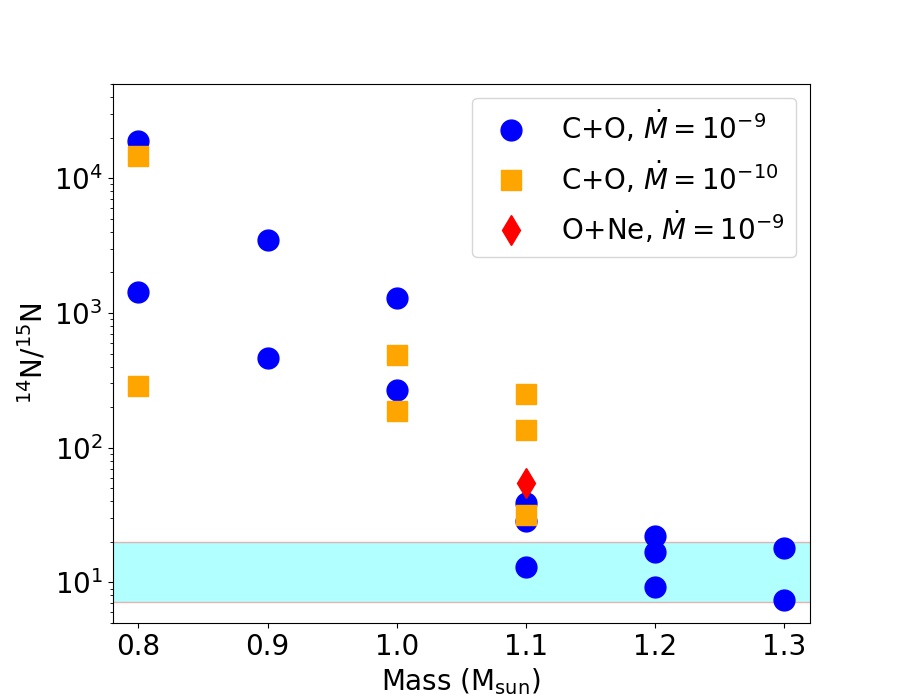}
    \includegraphics[width=5.75cm,trim=0.0in 0.0in 0.5in 0.6in,clip=True]{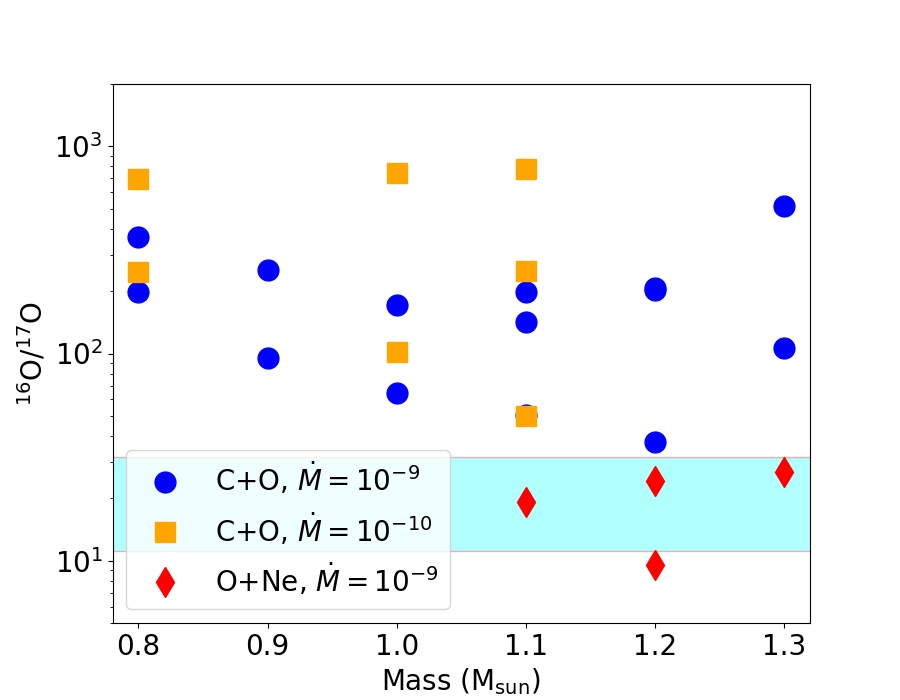}
    \caption{(left) $^{12}$C/$^{13}$C mass ratio for our nova models for different progenitor masses and accretion rates. The cyan box corresponds to the observational constraints taken from planetary nebula K4-47 \citep{Schmidt2018}. The orange box is the observed constraints taken from Nova Ophiuchus 2017 \citep{Joshi2017}. The vertical green box stands for the expected mass from UV spectra \citep{Mikolajewska2017}.
    (middle) $^{14}$C/$^{15}$N. 
    (right) $^{16}$O/$^{17}$O. 
    }
    \label{fig:nova_nucleo2}
\end{figure*}

We also examine the chemical abundance pattern of our nova models with focus on the lower mass elements C, N and O.
Most of them are produced in massive stars, but with a metallicity comparable to solar values.
Novae are often associated with super-solar production.
Here we examine how the production of these elements compare to the observational data.
We use measurements reported in \citet{Schmidt2018} who derived the chemical abundances from the planetary nebula K4-47.
The low $^{12}$C/$^{13}$C, $^{14}$N/$^{15}$N and $^{16}$O/$^{17}$O ratios highly deviate from the solar value, implying that standard massive star explosions are unlikely to be the origin.
Their ratio represents how the proton capture competes with the CNO cycle.
Another tight measurement from Nova Oph 2017 is given in \citet{Joshi2017} which also shows a very low $^{12}$C/$^{13}$C ratio of $1.6 \pm 0.3$. 
In Figure\,\ref{fig:nova_nucleo2} we show the three ratios of our models as a function of WD progenitor mass in comparison to these measurements.
%

There are subtle trends in the models albeit the large fluctuations among models.
For $^{12}$C/$^{13}$C, the mass ratio increases with the progenitor mass.
ONe WD models have a higher mass ratio than the CO WD models for high WD masses. 
For $^{14}$N/$^{15}$N, the mass ratio clearly decreases with the progenitor mass.
The ONe WDs show distinctively lower mass ratios.
There is no clear trend in the $^{16}$O/$^{17}$O mass ratio against the progenitor mass and ONe WDs again have the lowest value among all models.
The clustering of the data points suggests two trends to classify the CO and ONe WDs.
ONe WDs tend to have lower $^{14}$N/$^{15}$N ($< 100$) and lower $^{16}$O/$^{17}$O ($< 50$) ratios.
These thresholds could serve as an indicator of the WD composition.

For the observational data from \citet{Schmidt2018}, each observed isotope ratio suggests different mass ranges or WD types from our models.
Given the $^{12}$C/$^{13}$C-values from our models, progenitor masses of 1.0--1.1\,$M_{\odot}$ (or $\sim 0.8\,M_{\odot}$) are suggested.
From the ratio $^{14}$N/$^{15}$N as well as $^{16}$O/$^{17}$O, WD masses with 1.1--1.3\,$M_{\odot}$ are favored.
However, the overlap region around $1.1\,M_{\odot}$ would either suggest a low accretion rate on a CO WD ($^{12}$C/$^{13}$C), high accretion rate on a CO WD ($^{14}$N/$^{15}$N) or an ONe WD ($^{16}$O/$^{17}$O).
The displacement of the models with the data suggests that further input physics is required in matching all constraints at the same time, and it will be an interesting future project for an extensive exploration.

For the observational data from \citet{Joshi2017}, our models show that CO WDs with a mass 1.0--1.1\,$M_{\odot}$ or a $1.1\,M_{\odot}$ ONe WDs can approach the narrow $^{12}$C/$^{13}$C mass ratio observed in this nova.
The results are consistent with the spectrographic data in the UV band which also suggested that the observed WD has a mass of 1.0--1.1 $M_{\odot}$ \citep{Mikolajewska2017}.
In their work, the ONe WD is excluded. Our CO WD models with a low $f_{\rm CO}$ can fit the mass and the mass ratio simultaneously.
%

\section{Gamma-Ray Radiative Transfer}\label{sec:radtran}
After obtaining the kinematics of the ejecta from Section\,\ref{sec:evol} and the chemical abundance from Section\,\ref{sec:nucleo}, we pass the information to our Monte-Carlo radiative transfer code and compute the corresponding $\gamma$-ray spectrum.
From the spectrum we derive the $\gamma$-ray luminosity by aggregating the escaped photon packets.
To construct the spectra, we use logarithmic bins for photon energies below 508\,keV and above 514\,keV, up to 3\,MeV.

\subsection{Nova Gamma-Ray Spectra}\label{sec:gamma_spectra}

\begin{figure*}
    \centering
    \includegraphics[width=8cm,trim=0.0in 0.0in 0.5in 0.9in,clip=True]{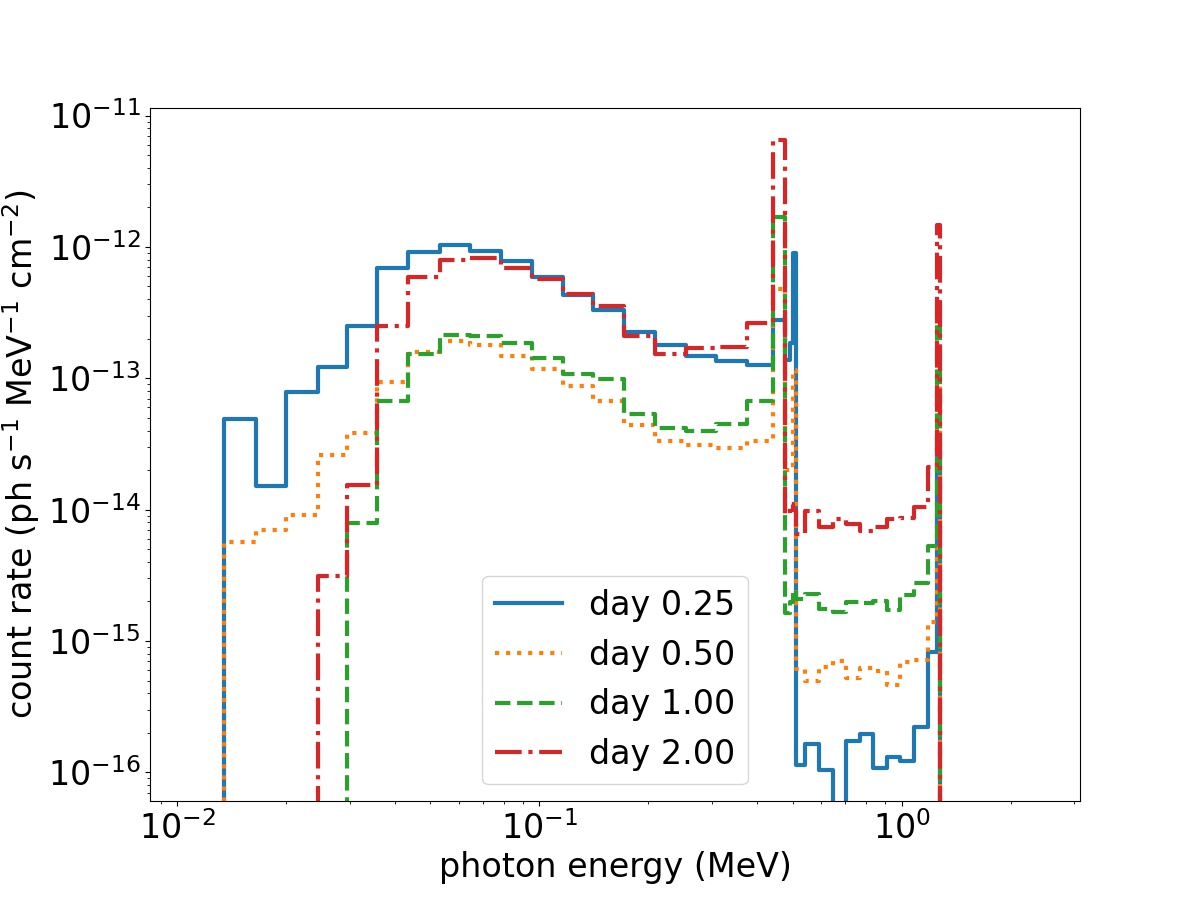}
    \includegraphics[width=8cm,trim=0.0in 0.0in 0.5in 0.9in,clip=True]{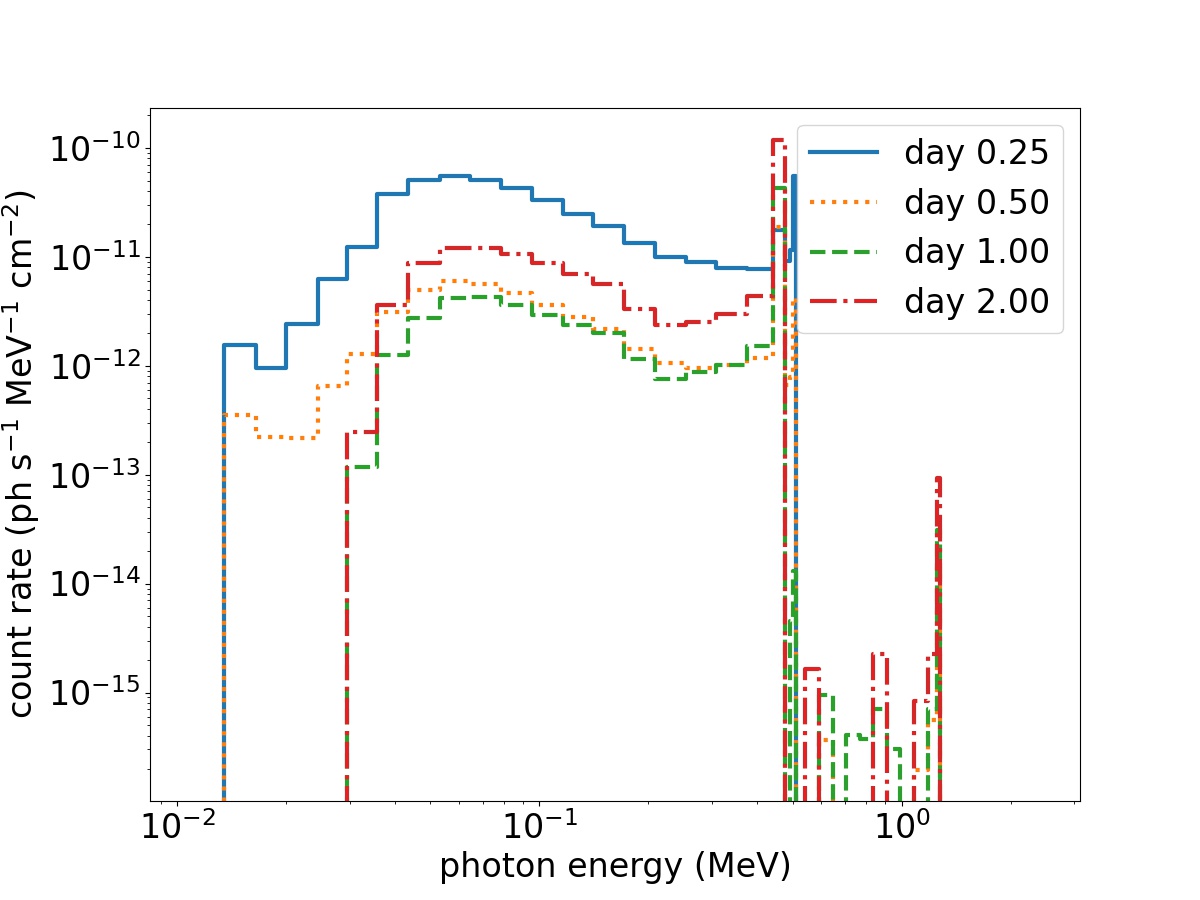}
    \includegraphics[width=8cm,trim=0.0in 0.0in 0.5in 0.9in,clip=True]{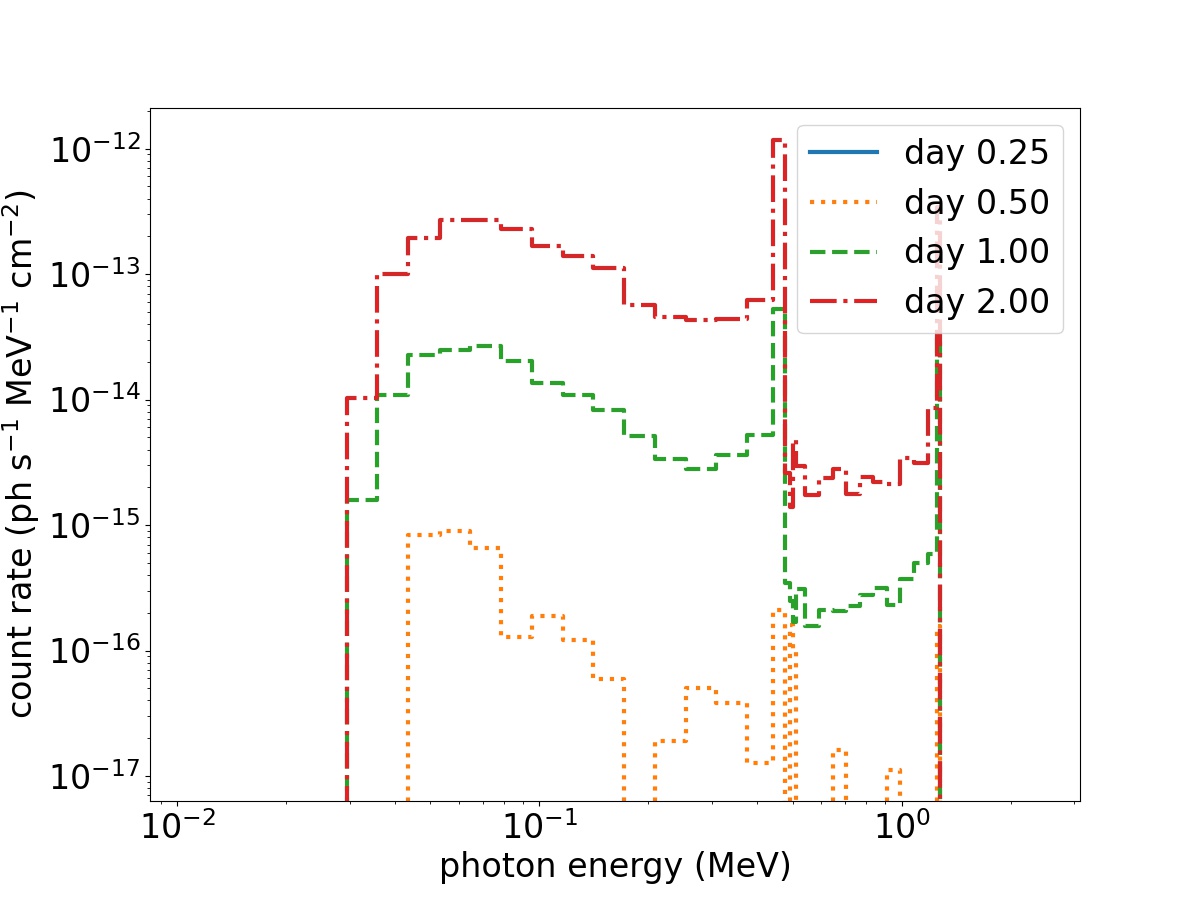}
    \includegraphics[width=8cm,trim=0.0in 0.0in 0.5in 0.9in,clip=True]{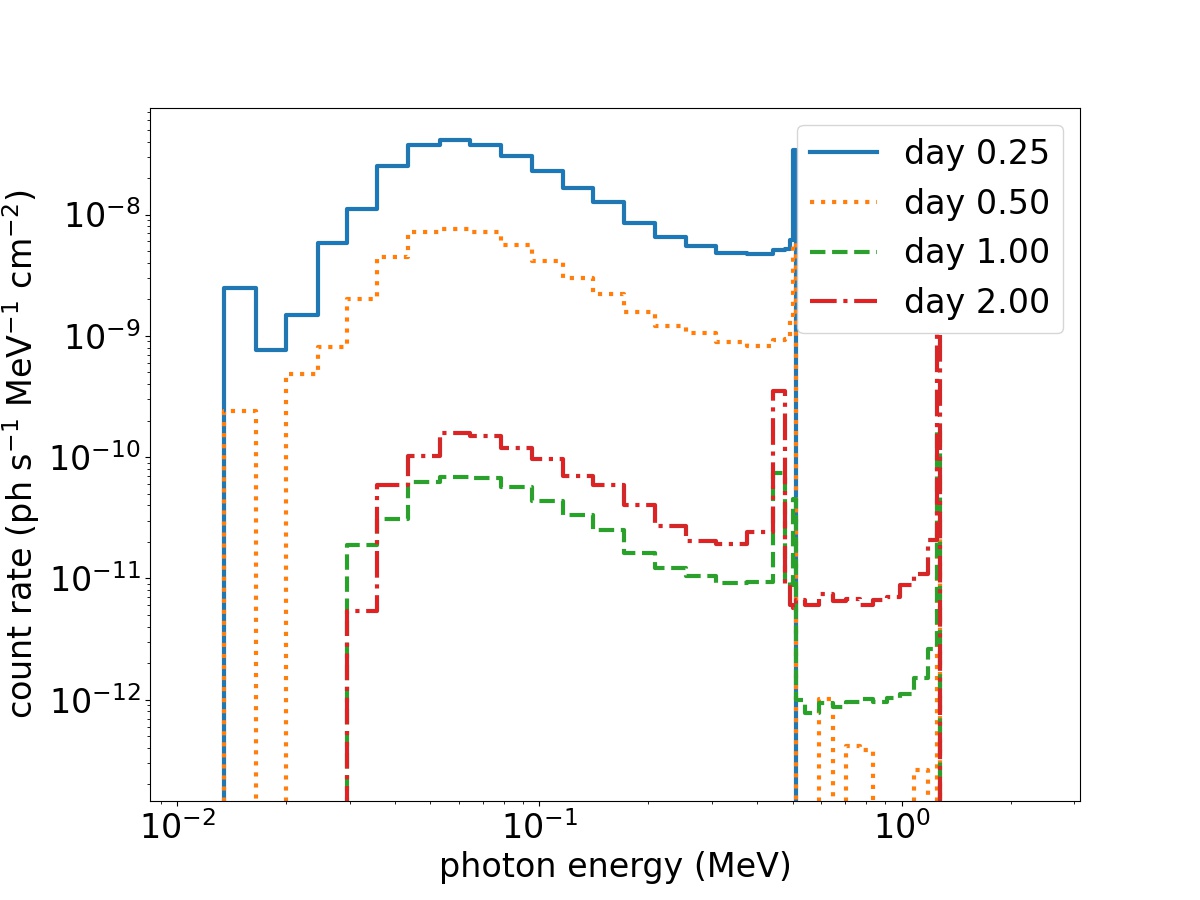}
    \caption{(top left) Gamma-ray spectra for the CO WD model with $M = 1.20\,M_{\odot}$, $\dot{M} = 10^{-9}\,M_{\odot}\,\mrm{yr^{-1}}$ and $f_{\rm CO} = 0.5$ at Day 0.25 (blue circle), 0.50 (orange cross), 1.00 (green square) and 2.00 (red triangle). (top right) CO WD model with $M = 0.8\,M_{\odot}$, $\dot{M} = 10^{-9}\,M_{\odot}\,\mrm{yr^{-1}}$ and $f_{\rm CO} = 0.5$. (bottom left) CO WD model with $M = 1.20\,M_{\odot}$, $\dot{M} = 10^{-9}\,M_{\odot}\,\mrm{yr^{-1}}$ and $f_{\rm CO} = 0.1$. (bottom right) ONe WD model with $M = 1.20\,M_{\odot}$, $\dot{M} = 10^{-9}\,M_{\odot}\,\mrm{yr^{-1}}$ and $f_{\rm CO} = 0.5$. The nova event is assumed to be at the distance of 1 kpc. 
    }
    \label{fig:star_spectra}
\end{figure*}

In Figure\,\ref{fig:star_spectra} we show the $\gamma$-ray spectra snapshots for different models at Day 0.25, 0.50, 1.00 and 2.00.
In all spectra, we observe that the low energy part $30$--$200$\,keV is dominated by the back-scattering of higher energy photons. 
Up to Day 0.50, the 511\,keV line from the $\beta^+$-ecays of \nuc{N}{13} and \nuc{F}{18} is dominating the spectra.
%
%
Starting at Day 1.00, the 478\,keV line from $^{7}$Be also emerges, and becomes important for later time as well due to its longer lifetime.
After Day 1.00, the 511\,keV line almost vanishes due to the exhaustion of $^{13}$N and $^{18}$F.
The spectral shape is almost unchanged for later times except that the absolute magnitude is decreasing with time.
There is an inversion in the CO WD models where the absolute magnitude of the spectra is greater at Day 1.00 than at Day 2.00.
This is because $^{7}$Be and $^{22}$Na, which support the later $\gamma$-ray luminosity, locate at the inner ejecta so that it takes a longer time for the photosphere to recess and expose these isotopes.

We find that the $\gamma$-ray luminosity is larger for ONe WD models compared to CO WD models.
ONe WD models have a higher accretion and outburst mass because ONe-rich matter requires higher masses to incinerate before the TNR to take place.
As a result, it generates more radioactive isotopes and hence stronger $\gamma$-ray emission.

A higher progenitor mass is beneficial for the emission of $\gamma$-rays.
This is because the maximum temperature reached during TNR is higher when the mass is higher.
The higher temperature allows more rapid nuclear reactions as well as higher thermal pressure, which helps to expel the matter.
Even though the progenitor WD is more compact for a higher mass WD, the overall burning is still stronger.

Finally, the $\gamma$-ray signal is sensitive to the mixing ratio $f_{\rm CO}$ of the CO (and without loss of generality ONe) matter.
Because the presence of CO matter facilitates the outburst process not only by its interval, but also its tendency to burn.
A high $f_{\rm CO}$ allows the outburst of the envelope with a smaller ejecta mass and a higher velocity which facilitates the transport of $\gamma$-rays.

\subsection{Nova Gamma-Ray Light Curves}\label{sec:lightcurves}

\begin{figure*}
    \centering
    \includegraphics[width=8cm,trim=0.0in 0.0in 0.5in 0.9in,clip=True]{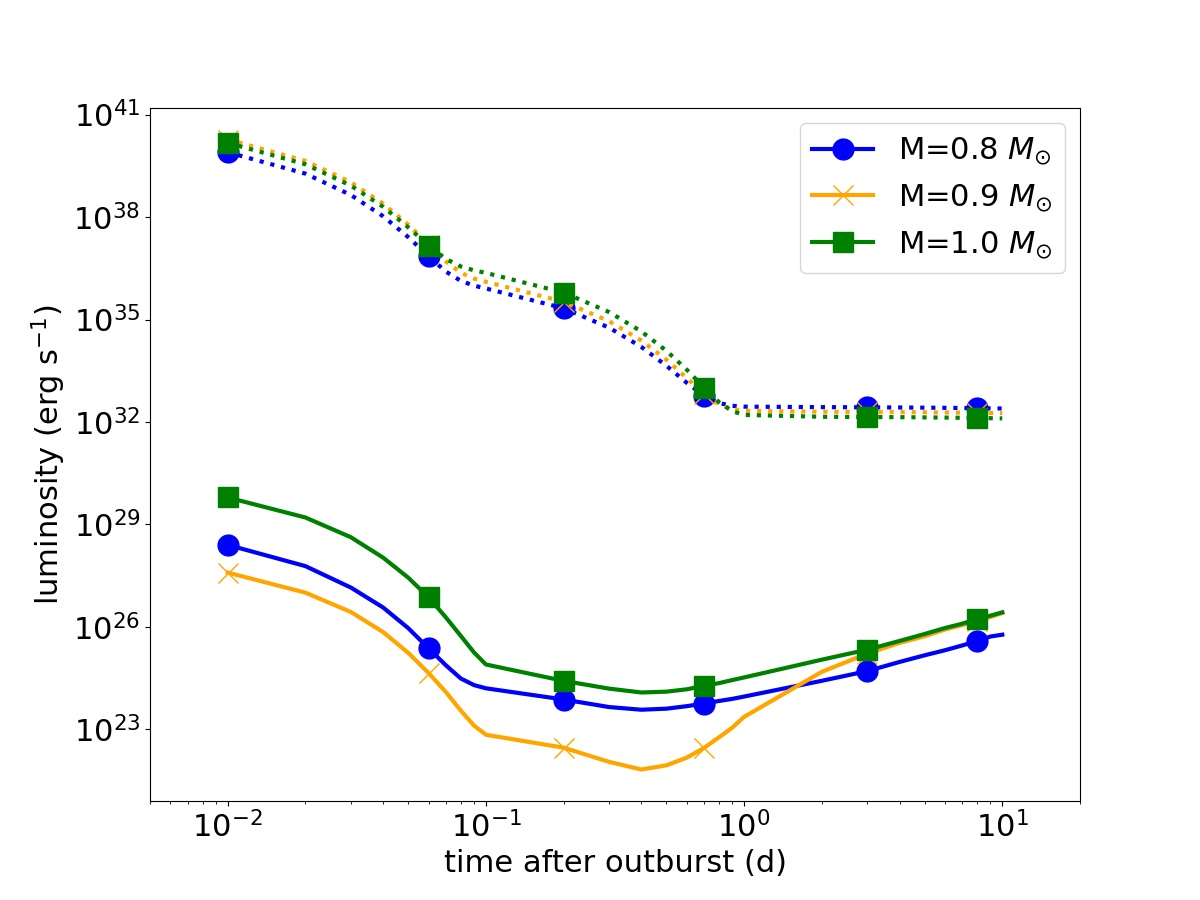}
    \includegraphics[width=8cm,trim=0.0in 0.0in 0.5in 0.9in,clip=True]{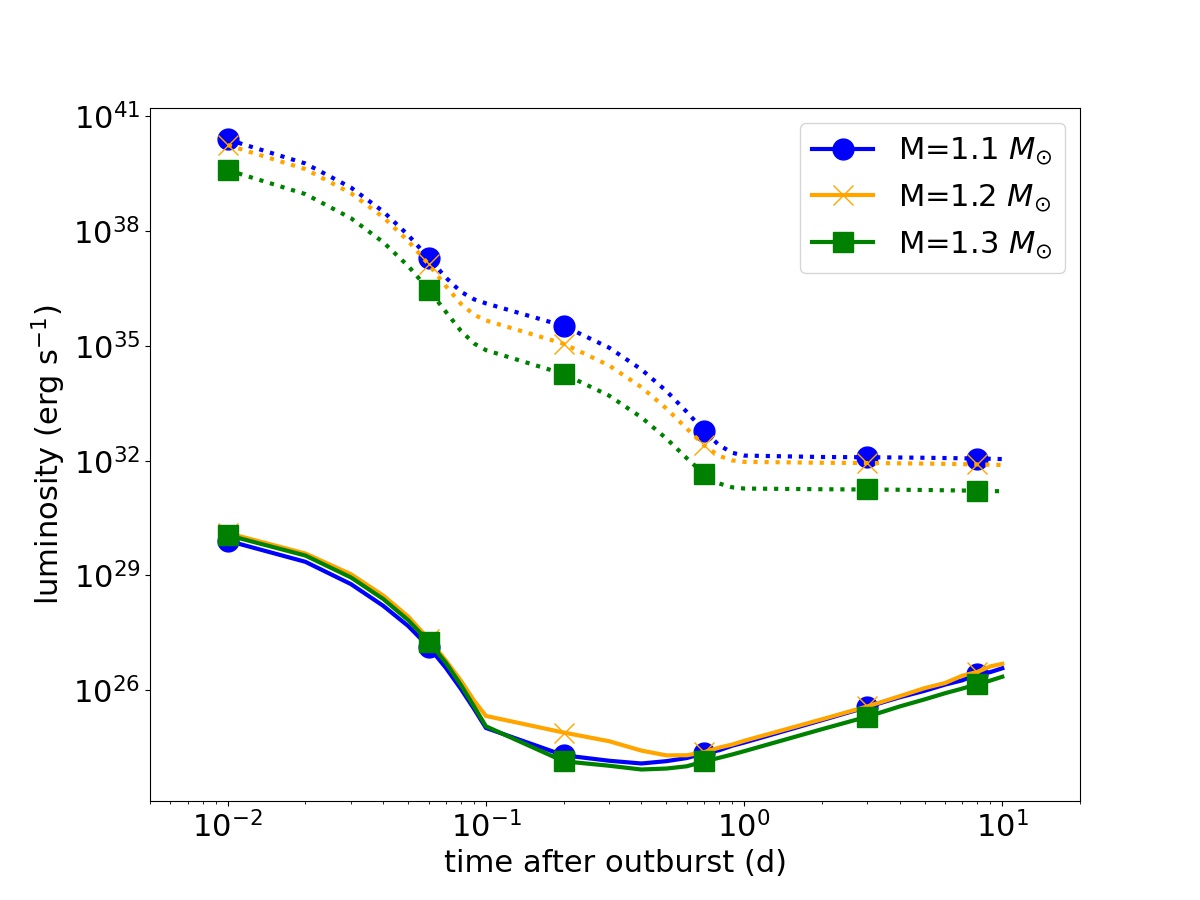}
    \includegraphics[width=8cm,trim=0.0in 0.0in 0.5in 0.9in,clip=True]{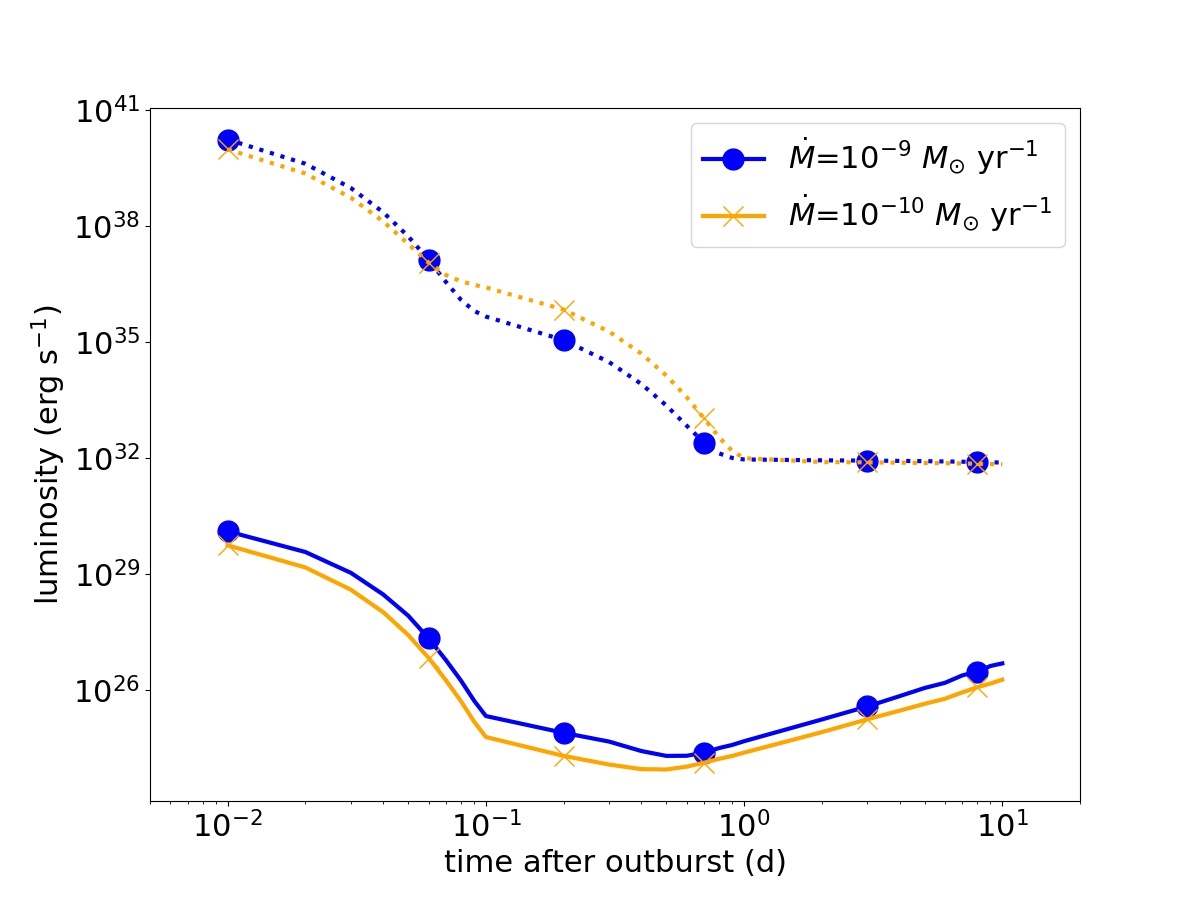}
    \includegraphics[width=8cm,trim=0.0in 0.0in 0.5in 0.9in,clip=True]{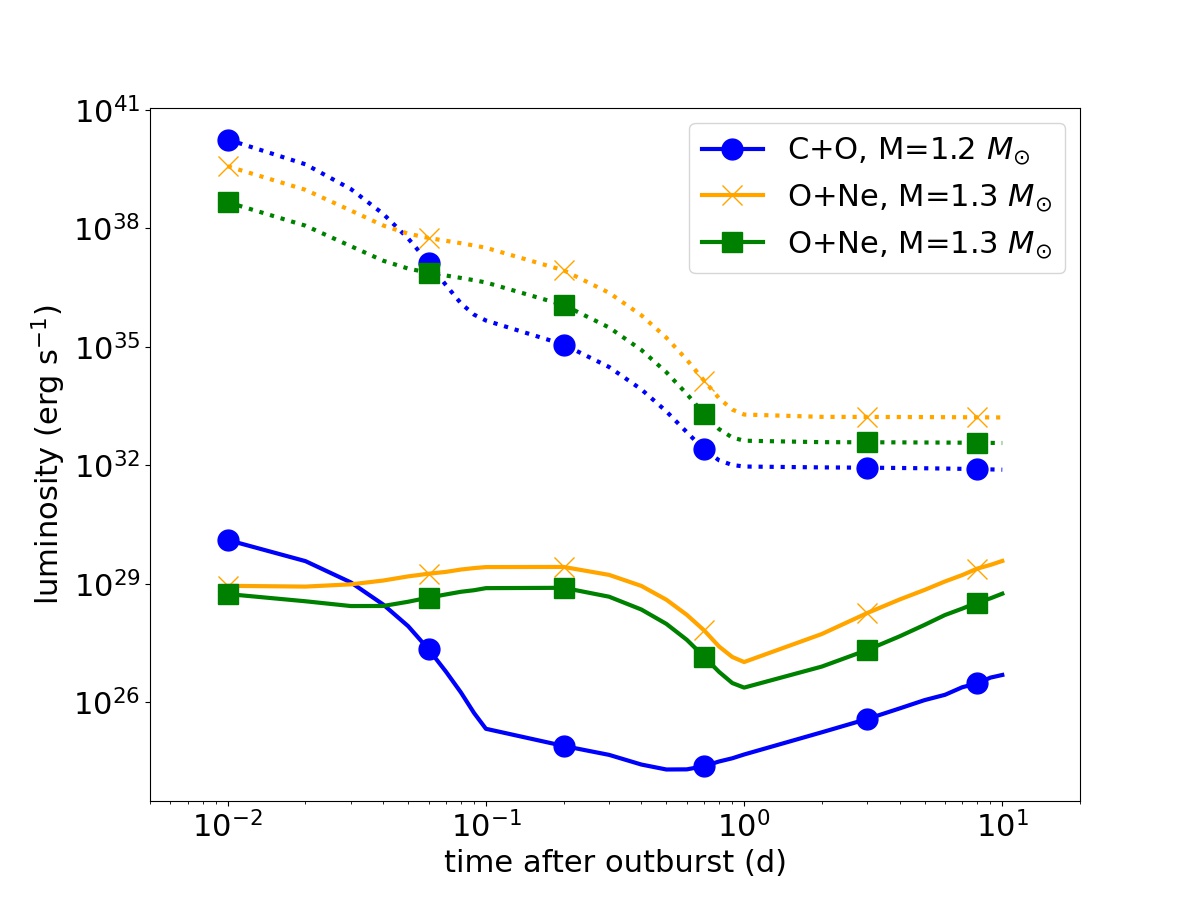}
    \caption{(top left) $\gamma$-ray light curves of CO WD nova models with $f_{\rm CO}=0.5$, $\dot{M}=10^{-9}\,M_{\odot}$\,yr$^{-1}$ and $M = $0.8 (blue circle), 0.9 (orange cross) and 1.0\,$M_{\odot}$ (green square), respectively. The total radioactive decay luminosity (dotted line) is included for reference. 
    (top right) CO WD models with 1.1 (blue circle), 1.2 (orange cross) and 1.3 $M_{\odot}$ (green square), respectively. The light curve luminosity integrates the gamma spectra from 10 keV to 3 MeV.  
    (bottom left) CO WD models with $M = 1.2\,M_{\odot}$, $f_{\rm CO} = 0.5$ and $\dot{M}=10^{-9}\,M_{\odot}$\,yr$^{-1}$ (blue circle) and $\dot{M}=10^{-10}\,M_{\odot}$\,yr$^{-1}$ (orange cross). 
    (bottom right) ONe WD models with $M = 1.2$ (orange cross) and 1.3\,$M_{\odot}$ (green square). The CO WD model with $M = 1.2\,M_{\odot}$ is included for comparison 
    }
    \label{fig:star_LC}
\end{figure*}

By collecting the corresponding $\gamma$-ray luminosity in the energy band from 10 keV to 3 MeV, we show the light curve of representative models in Figure\,\ref{fig:star_LC}.
Unlike for type Ia supernovae where the luminosity is very low at early times, the $\gamma$-ray luminosity for novae is the highest at the beginning:
Because of the `exposed' $^{13}$N and $^{18}$F on the surface the direct formation and decay of Ps leads the $\gamma$-rays to readily escape from the star.
The luminosity reaches as high as $\sim 10^{28}\,\mrm{erg\,s^{-1}}$.
However, the luminosity quickly drops by five to six orders of magnitude soon after.
The recession of the photosphere is slow so that the escape of photon packets from the inner layers is very inefficient.
After Day 1.00 after the outburst, the $\gamma$-ray luminosity gradually increases and reaches an asymptotic value of $\sim 10^{25}\,\mrm{erg\,s^{-1}}$.
We find that the early emission is dominated by the 511\,keV line as has been found in previous studies.
\begin{figure*}
    \centering
    \includegraphics[width=8cm,trim=0.0in 0.0in 0.5in 0.9in,clip=True]{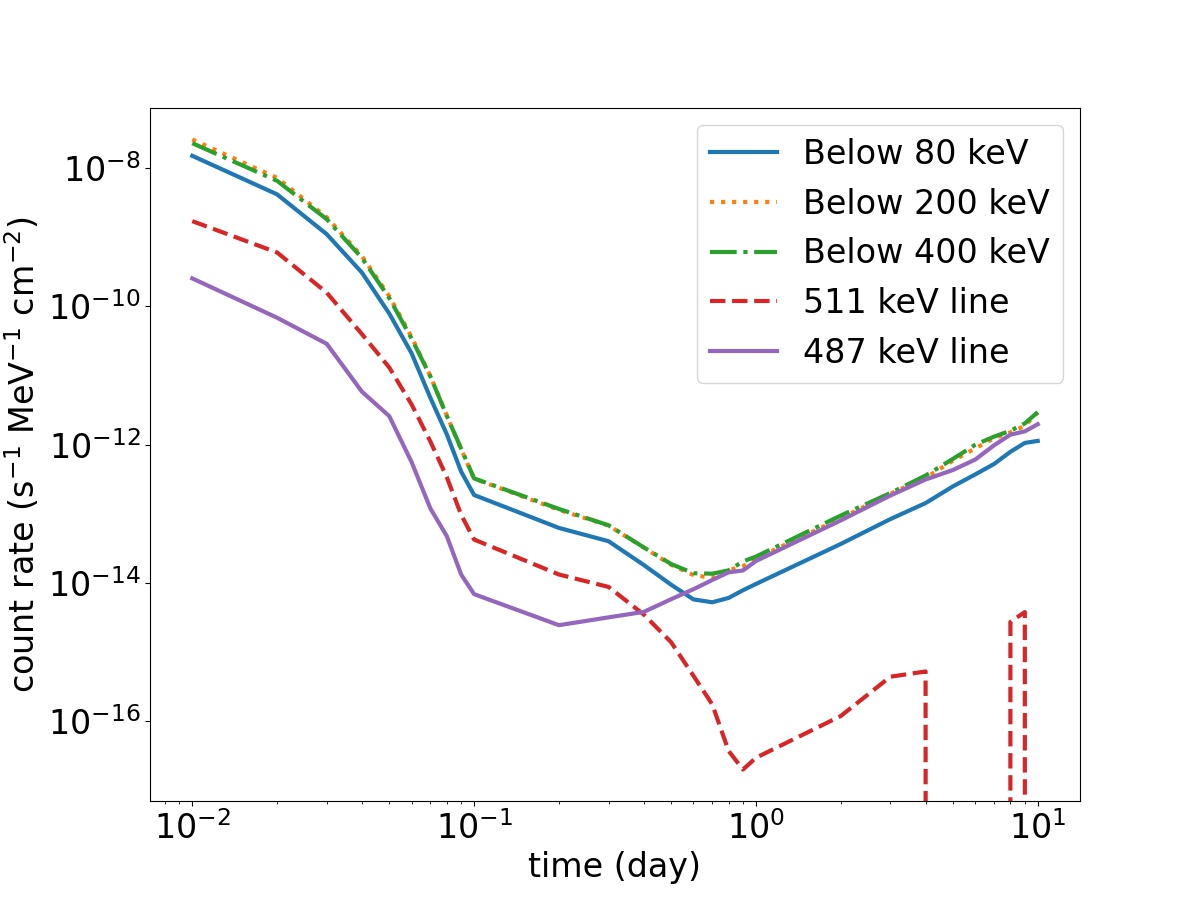}
    \includegraphics[width=8cm,trim=0.0in 0.0in 0.5in 0.9in,clip=True]{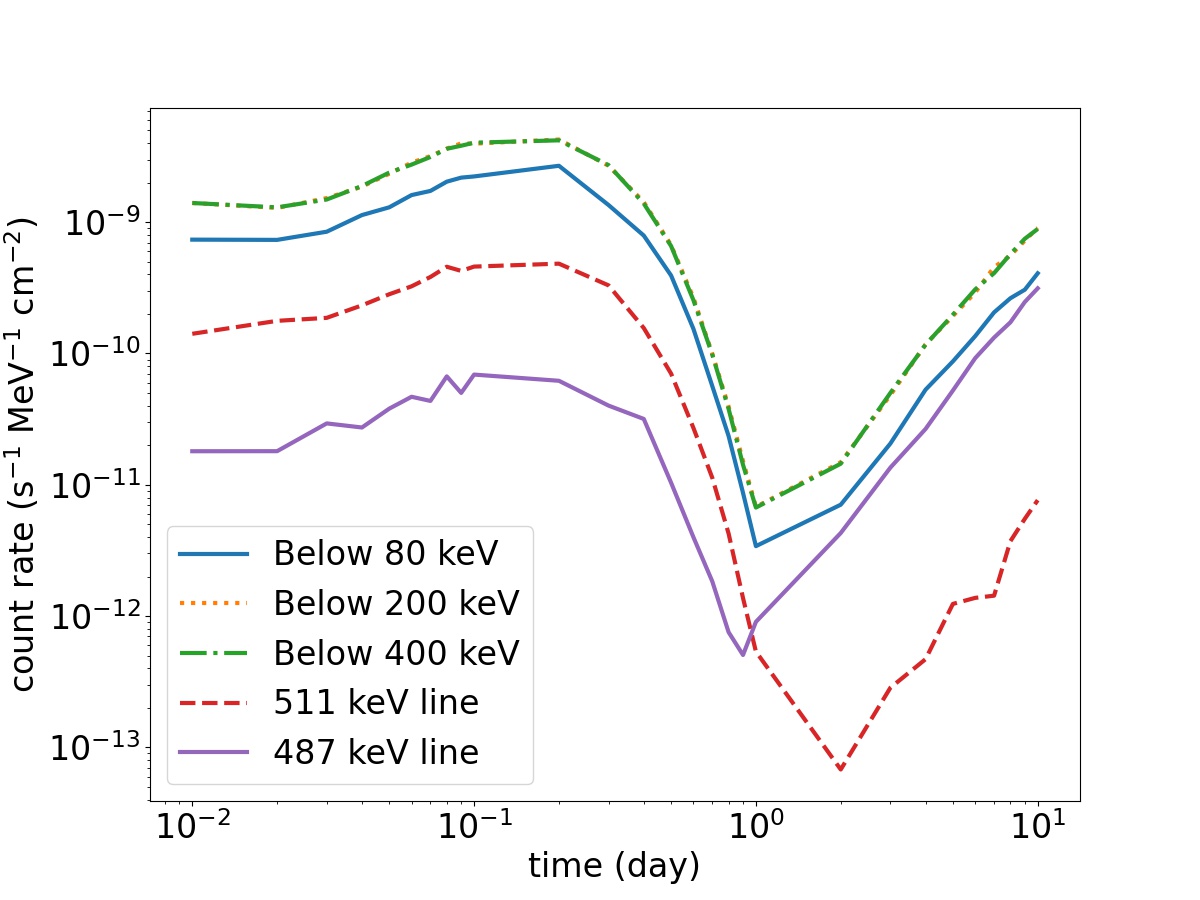}
    \caption{\textit{(left)} The light curve components for the CO WD model with $M = 1.20~M_{\odot}$, $\dot{M}=10^{-9}~M_{\odot}$ yr$^{-1}$ and $f_{\rm CO} = 0.5$ assuming the nova is at 1 kpc apart from the Earth. \textit{(right)} Same as the left panel but for the ONe WD model with the same configuration.}
    \label{fig:LC_components}
\end{figure*}

We examine the dependence of the $\gamma$-ray light curve on the WD progenitor:
A higher mass WD generates a higher $\gamma$-ray luminosity but it converges quickly for models with $M > 1.1\,M_{\odot}$.
The changes due to different $\dot{M}$ are small, as the light curve is primarily depending on the TNR strength and the synthesized radioactive isotopes (cf. previous sections).
The WD composition has a significant effect on the light curve shape.
CO WDs show a rapidly falling luminosity, which is different from the parabolic shape of light curves appearing in ONe WD models.
The parabolic shape implies that the lines related to early times are stronger in the ONe WD models, compared to CO WD models.

To further extract the $\gamma$-ray signals, we plot in Figure\,\ref{fig:LC_components} the components of our reference models assuming $M = 1.20\,M_{\odot}$, $\dot{M}=10^{-9}\,M_{\odot}$\,yr$^{-1}$ and $f_{\rm CO} = 0.5$ for the CO WD (left panel) and the ONe WD (right panel).
Most of the early time light curve power in the CO model comes from the scattering background below 80\,keV and from the 511\,keV line (here we define 511$\pm$6 keV as the line), albeit weaker than the scattering background.
At Day 1 and beyond, the decay of $^{7}$Be becomes dominant in the light curve.
In contrast, 511\,keV line becomes insignificant because most short-life radioactive isotopes $^{13}$N and $^{18}$F have decayed.
The ONe model shows a similar hierarchy in its luminosity components.
The difference in the chemical composition makes the early time ($<1$\,day) scattering background and 511\,keV line stronger for a longer time.

The 487\,keV line from $^{7}$Be decay surpasses the 511\,keV line after Day 1.
Beyond Day 10, as the ejecta become transparent in regions where they have been accelerated by thermal expansion, the escaped $\gamma$-ray luminosity approaches the total luminosity by radioactivity decay.
For example, the reference model takes about 30 days until the escaped $\gamma$-ray luminosity is 99\% of the total luminosity.

\section{Applications to Observations}\label{sec:application}
In general, our new models can be used to improve $\gamma$-ray searches from individual novae as well as to study the cumulative effect of the whole nova population in the Milky Way.
First we note that our flux estimates are at least four to eight orders of magnitude smaller within the first $\sim 30$ days after the explosion compared to previous calculations \citep[e.g.,][the first paper is called JH98 hereafter]{Jose1998_novae,GomezGomar1998,Hernanz2014_nova}.
This would make any detection prior to a month after the explosion impossible for a nova happening at any astrophysical distance for current instrumentation.
In Tab.\,\ref{tab:fluxes} we provide a list of integrated fluxes for characteristic times and energy bands, comparing the brightest model variants of a CO and an ONe WD with $1.2\,M_{\odot}$ each, as well as the CO WD simulation setup from JH98.
In particular the supposed `511\,keV-flash' hours after the outburst that had been suggested to show fluxes on the order of $10^{-3}$--$10^{-1}\,\mrm{ph\,cm^{-2}\,s^{-1}}$ would be reduced to $10^{-10}$--$10^{-9}\,\mrm{ph\,cm^{-2}\,s^{-1}}$ in our new model calculations.

\begin{figure*}
    \centering
    \includegraphics[width=8cm,trim=0.0in 0.0in 0.5in 0.9in,clip=True]{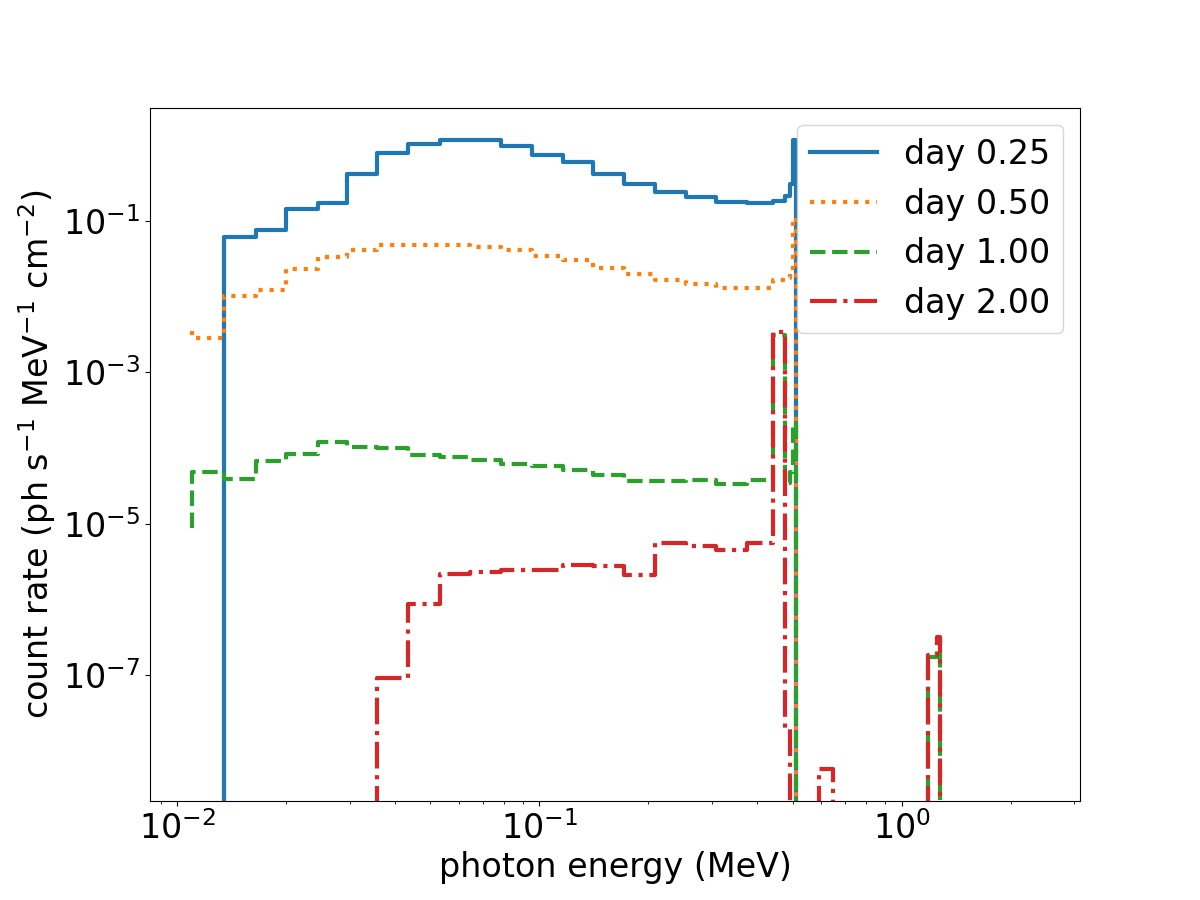}
    \includegraphics[width=8cm,trim=0.0in 0.0in 0.5in 0.9in,clip=True]{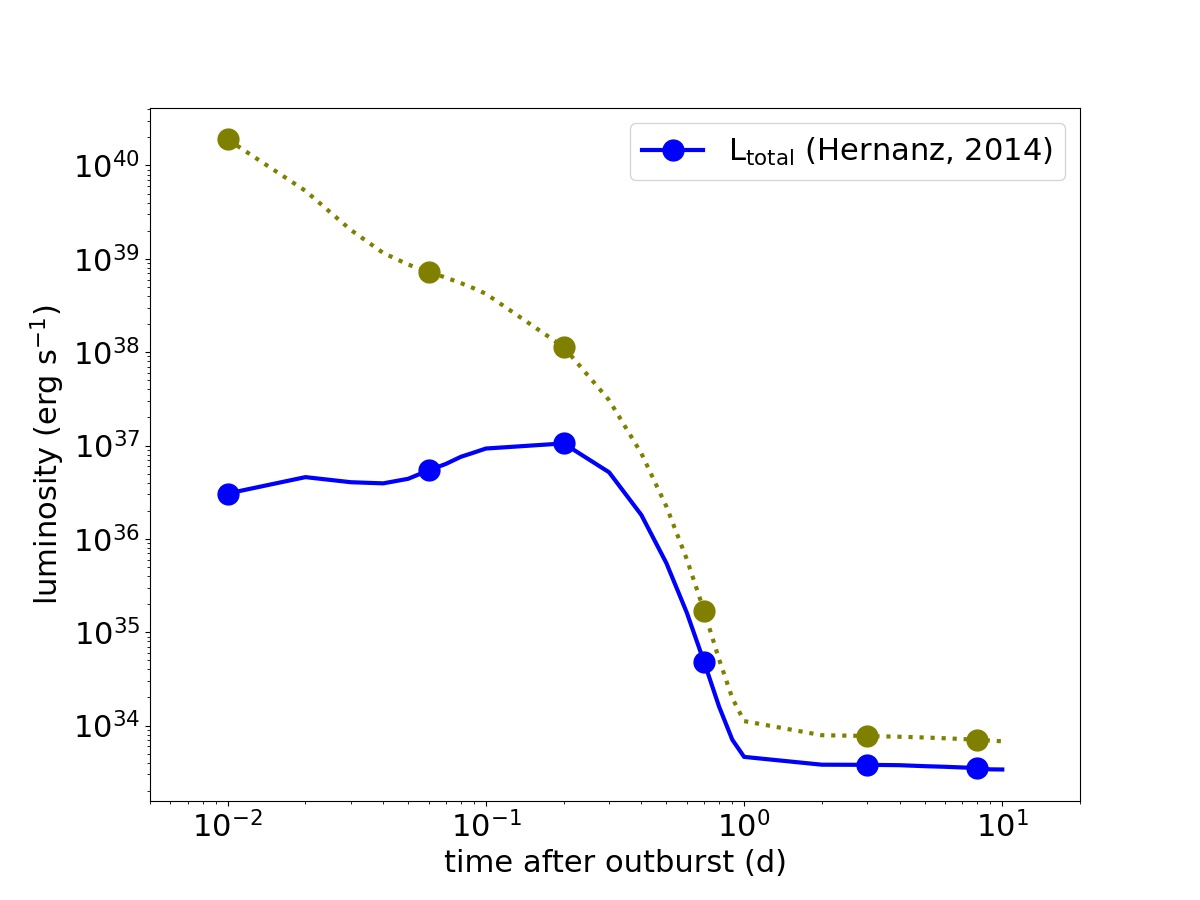}
    \caption{(left) Spectra of the nova model with $M = 1.20\,M_{\odot}$, $f_{\rm CO} = 0.5$ and $\dot{M} = 10^{-9}\,M_{\odot}$\,yr$^{-1}$ at Day 0.25, 0.50, 1.00 and 2.00 after the mass ejection using the configuration taken from \citet{Hernanz2014_nova}. 
    (right) Light curve (blue solid line) and the total luminosity (olive dotted line).
    }
    \label{fig:Hernanz_spectra}
\end{figure*}

\begin{table}
\caption{Comparison of the brightest models with the configurations $M_{\rm WD} = 1.2\,M_{\odot}$ and $f_{\rm CO} = 0.5$ for specific times and energy bands. The fluxes are given in units of $\mrm{ph\,cm^{-2}\,s^{-1}}$.}
\begin{center}
\begin{tabular}{c|c|ccc}
    Model & Time [d] & $<80$\,keV & $<200$\,keV & $511$\,keV \\
    \hline
    \multirow{3}{*}{CO} 
     & $0.01$ & 1.3e-08 & 2.6e-08 & 2.0e-09 \\
     & $0.20$ & 5.7e-14 & 1.2e-13 & 1.4e-14\\
     & $1.00$ & 8.1e-15 & 2.1e-14 & 3.0e-17\\
    \hline
    \multirow{3}{*}{ONe} 
     & $0.01$ & 6.7e-10 & 1.5e-09 & 1.5e-10\\
     & $0.20$ & 2.1e-09 & 4.4e-09 & 4.7e-10\\
     & $1.00$ & 2.9e-12 & 6.7e-12 & 5.4e-13\\
    \hline
    \multirow{3}{*}{JH98} 
     & $0.01$ & 5.8e-02 & 1.2e-01 & 1.3e-02\\
     & $0.20$ & 1.7e-00 & 3.5e-00 & 3.6e-01\\
     & $1.00$ & 6.0e-06 & 1.1e-05 & 2.6e-06\\
    \hline
\end{tabular}
\end{center}
\label{tab:fluxes}
\end{table}

Such an extremely reduced flux in the entire 0.02--2\,MeV energy band has also an impact on the interpretability of diffuse emission measurements of the 478 and 1275\,keV lines:
For example, \citet{Siegert2021_BHMnovae} performed the latest analysis of diffuse emission from novae in the Milky Way.
Because the early-time light curves from \nuc{Be}{7} and \nuc{Na}{22} have typically been assumed to follow an exponential radioactive decay with maximum luminosity around days 4--6 after the explosion \citep{Hernanz2014_nova}, the ejecta masses are wrongly calibrated in such observations.
Extrapolating from years after the explosion with pure radioactive decay will lead to higher ejecta mass estimates than what might actually be the case, especially if the peak $\gamma$-ray luminosity occurs only several weeks after the explosion.
Therefore, with our new models, the cumulative effect of novae in the Milky Way can be estimated with higher accuracy in a population synthesis model, and then fitted via a Bayesian Hierarchical Model similar to one performed by \citet{Siegert2021_BHMnovae}.

Another application concerns the positron annihilation flash shortly after the explosion:
Instead of using only the 511\,keV line or individual energy bands, as has been done in previous retrospective studies \citep[e.g.,][]{Harris1999_novaflash,Harris2000_novaflash,Hernanz2000_novaflash,Smith04,Skinner2008_nova511_retrospective}, MeV data analysis is severely more sensitive to expected signals if the entire time-variable spectrum is used.
Soft $\gamma$-ray analysis has been recently progressing into the regime of full forward-modelling, i.e. taking into account the full response and position of the source in a single step -- without assuming generic spectral shapes such as power-laws.
Because our models are time resolved, and in addition would include several other parameters such as the mixing fraction, accretion rate, or WD mass, it is possible to perform both, a directed retrospective search for known novae, and a blind search for unknown objects.
Because the optical maximum from classical novae occurs 2--10 days after the explosion \citep{GomezGomar1998}, the expected 511\,keV flash has already passed.
However there is a chance that individual $\gamma$-ray instruments observed the nova by chance, or that an all-sky monitor showed enhanced fluxes but which could not be identified.
In a directed retrospective search for MeV emission from past nova outbursts, the expected time frame is revisited in data archives and investigated for coincident emission.
With our new time-resolved models and the ability to perform simultaneous fits in time and energy \citep[for example with the 3ML framework,][]{Vianello2015_3ML}, a directed retrospective search for $\gamma$-ray flashes from novae can be conducted.
Ideal instruments currently in space would be Swift/BAT, Fermi/GBM, and ISGRI and SPI on INTEGRAL.
With their large fields of views the chances are high that individual novae happened close enough for them to be detectable.
In fact, the number of novae that should have been detected by SPI within the last 19 years, based on previous flux estimates from JH98 of the 511\,keV line alone, is between 1 and 40.
Since no nova or 511\,keV flash has been reported from SPI, our results of an extremely reduced flux seem to agree with these non-detections.

A general retrospective search for MeV emission of unknown novae in the Milky Way is now also possible:
The nova rate in the Milky Way is $50 \pm 25\,\mrm{yr^{-1}}$ \citep{Shafter2017_novarate}, however only 10--30\,\% are detected at UVOIR wavelengths each year.
Such a search would be very time consuming because in addition to the time and energy domain also the position would need to be determined.
Because the highest fluxes are still expected within the first 2--3 hours after the explosion, several overlapping time frames of observations will be required to obtain reasonable baselines to search for unexplained emission in addition to known sources.
Thanks to developments in Fermi/GBM background modelling \citep{Biltzinger2020} that allows to search also for longer signals, as well as SPI (Biltzinger et al. 2022, in prep.) and Swift/BAT \citep{DeLaunay2021_BATGUANO} data analysis, such a blind search might be worthwhile.

Given the enhanced line and continuum sensitivity of the new NASA SMEX mission COSI \citep{Tomsick2019_COSI}, on the order of a few novae should be detectable within its nominal mission time of two years.
Our models suggest that the brightest MeV $\gamma$-ray emission only occurs several weeks after the nova explosion and not days.
Because COSI is observing the full sky within 24 hours, an observation strategy is not required here.
However, this shifted maximum in the 478\,keV line, for example, should be considered for targeting observatories such as INTEGRAL whenever a nearby nova will trigger dedicated observations.

\begin{figure*}
    \centering
    \includegraphics[width=\textwidth,trim=0.4in 0.3in 0.3in 0.3in]{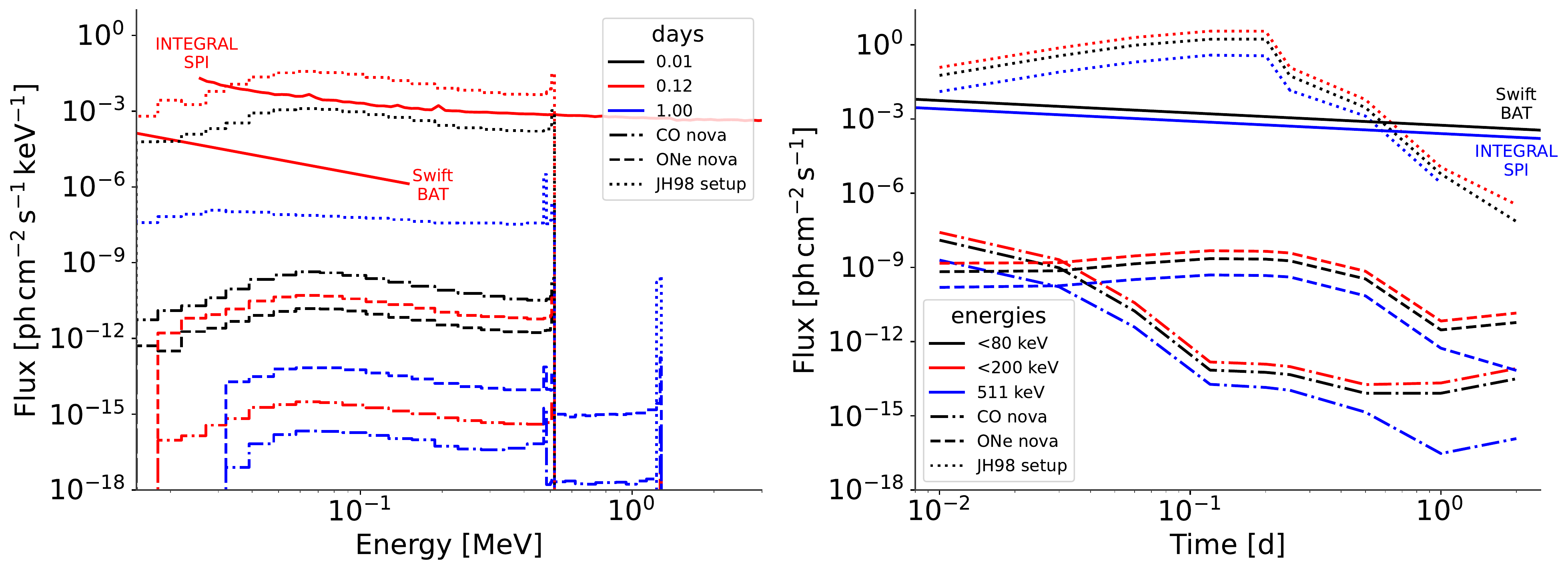}
    \caption{Comparison of brightest emission models of this work (CO, dash-dotted, and ONe dashed), and calculations with simulation setups in the literature (JH98, dotted). The distance to the nova is assumed to be 1\,kpc. For comparison, the sensitivities for INTEGRAL/SPI and Swift/BAT are given for day 0.12 (red, left), and the band containing photons with energies $< 80$\,keV (black, right) and of the 511\,keV line (blue, right).}
    \label{fig:flux_comparisons}
\end{figure*}

\section{Discussion}\label{sec:discuss}
\subsection{Comparison with Literature Models}\label{sec:literature_comparison}
\subsubsection{\citet{GomezGomar1998}}\label{sec:gomez-gomar}
In JH98 radiative transfer of selected CO and ONe nova models are presented.
They used a one-dimensional hydrodynamics model to compute the deposition of matter and their outburst with a large 100-isotope networks up to $^{40}$Ca.
The used accretion rate was $\dot{M} = 2\times10^{-10}\,M_{\odot}$\,yr$^{-1}$ with $f_{\rm CO} = 0.5$.
To check that our code can reproduce qualitatively similar results, we calculate an artificial model by setting the expansion velocity to $3000$\,km\,s$^{-1}$ with the radioactive isotope masses in the same order of magnitude as theirs.
In Figure\,\ref{fig:Hernanz_spectra} we show the spectra at selected times and the light curve from 0.01--10 days.

By comparing the energetics and thermodynamics of this and our models, our model shows a comparable production of $^{7}$Be, $^{13}$N and $^{22}$Na.
The $^{18}$F abundance is lower by one to two orders of magnitude.
Our ejecta velocity is also smaller in general.
Their models show a rapid expansion of about\,3000 km\,s$^{-1}$, while our models show a very slow expansion right after the TNR.
The ejecta reach about 300\,km\,s$^{-1}$ only at $\sim$ Day 10.
The peak temperature reached in their models is higher than our models by 0.2 in $\log_{10}$ scale.

Our spectra show agreeing results with theirs:
at the beginning, a flux of $\sim 10^{-2}\,\mrm{ph\,cm^{-2}\,s^{-1}}$ is observed in the 511\,keV line.
At Day 1 and beyond, the 511\,keV line becomes subtle while the 478\,keV line becomes prominent.
Meanwhile, the weak but observable 1275\,keV line emerges as well.
In Fig.\,\ref{fig:flux_comparisons} we show the spectra of our brightest models as well as the resulting spectrum from JH98 for three selected times, 15\,min, 3\,h, and 1\,d after the explosion, and compare them to instrument sensitivities.
Within a distance of 1\,kpc, none of the new models would result in measurable emission.

Our light curves show also similar results compared to theirs:
the early light curve is rapidly falling and then reaches an asymptotic value of about $10^{34}$\,erg\,s$^{-1}$ after Day 1.
The artificially high velocity makes the ejecta transparent at much earlier times, so that the escaping $\gamma$-rays approach the total radiative decay power much faster than in our models.
A factor of two is observed between total and escaped $\gamma$-rays.
Also shown in Fig.\,\ref{fig:flux_comparisons} is a comparison of the lightcurves between our models and previous calculations.
Again it is evident that current instrumentation would not be able to observe the short-duration nova flash.
%

\subsection{Caveats}\label{sec:caveats}
The mixing process is known to be complex for novae as the mixing changes the composition of the material \citep{Fujimoto1988}, which affects the recurrence time, the ignition temperature and the corresponding nuclear reactions \citep{Shen2009, Denissenkov2013}.
The stellar evolution model used here assumes spherical symmetry, while the mixing involving Kelvin-Helmholtz and baroclinic instabilities naturally involves modeling with two- or three-dimensional simulations \citep[e.g.,][]{Casanova2016, Casanova2018}.
Thus it is unclear whether the constant mixing and the mixing rate used in this work is completely agreeing with the actual value.
The combination of using multi-dimensional simulations to extract the mixing parameters with application to spherical symmetric models \citep{Jose2020} can provide a better estimate how the mixing takes place.

The production of $^{7}$Be has been in tension as theoretical models produce about $10^{-5}$ in mass fraction \citep{Chugai2020} which is an order of magnitude lower than the averaged observational value $\sim 10^{-4}$ \citep{Molaro2020}.
This might however be an observational bias toward extremely large values.
Our models, where $^{7}$Be shows a mass fraction of about $10^{-5}$ agree with the consensus of theoretical models but require further input physics to match the observational data for these extreme measurements.
Recent works searching for optimized parameters suggest that initial composition including $^4$He still plays a role to the exact abundance pattern \citep{Chugai2020, Denissenkov2021}, but cannot completely resolve the tension.
Since the $^{7}$Be decay has a half life $\sim 54$ days, the exact amount will largely contribute to the later $\gamma$-ray spectra when the ejecta become transparent.
The tension also suggests that more careful treatment in the pre-outburst evolution is necessary. 
Similar to type Ia supernovae, a single detection in the MeV $\gamma$-ray band will not necessarily clarify this conundrum because there is a variety of parameters that will influence the resulting spectrum and in turn our understanding of classical novae.

Our models predict in general the ejecta with a velocity much lower than previous works \citep[e.g.,][]{Jose2001}.
Differences can be seen from Figs.\,\ref{fig:models_Mej} and \ref{fig:models_Tmax}, where their models show a higher $T_{\rm max}$ and $M_{\rm ej}$.
A major difference comes from the choice of radiative opacity.
They have used the table from \citet{Iben1975} while ours use \citep[][I93]{Iglesias1993}.
As shown in the two figures, their models show a large reduction of $T_{\rm max}$ and $M_{\rm ej}$ when the I93 opacity table is used.
The more recent opacity table predicts softer matter, which can be compressed and heated more easily, leading to a TNR faster and more frequent.
These effects impact to the $\gamma$-ray transport, so that it takes more time for the ejecta to become optically thin and consequently for the $\gamma$-rays to escape.

\subsection{Conclusion}\label{sec:conclusion}
In this work we have used the stellar evolution code MESA for computing an array of CO and ONe nova models for a given WD mass $M_{\rm WD}$ from 0.8--1.3\,$M_{\odot}$, mass accretion rate $\dot{M} = 10^{-10}$--$10^{-9}\,M_{\odot}$\,yr$^{-1}$ and mixing ratio $f_{\rm CO} = 0.1$--$ 0.5$.
We study the parameter dependence of the nova recurrence, thermodynamics, and chemical abundance pattern.
We find that some of the nova models can fit to the extremely low values of $^{12}$C/$^{13}$C, $^{14}$N/$^{15}$N, $^{15}$O/$^{16}$O observed in planetary nebulae and novae.

We then carry out Monte-Carlo radiative transfer simulations to study the soft $\gamma$-ray spectra and corresponding light curves.
A major result of this work is that novae take a longer time ($\sim 5$--$10$ days) before the ejecta reach a high velocity ($\sim 100$--$300$\,km\,s$^{-1}$) for the matter to become transparent to $\gamma$-rays.
As a result, most of the 30--500\,keV emission and in particular the 511\,keV line coming from the short-lived radioactive isotopes such as $^{13}$N and $^{18}$F is heavily suppressed. 

In particular we find that the previously expected 511\,keV line flash with fluxes on the order of $10^{-3}$--$10^{-1}\,\mrm{ph\,cm^{-2}\,s^{-1}}$ within the first few hours after the explosion is reduced to $10^{-16}$--$10^{-9}\,\mrm{ph\,cm^{-2}\,s^{-1}}$, depending on the model configuration.
The high end of these flux expectations are only reached for high mixing values $f_{\rm CO} \sim 0.5$.
With current instrument capabilities from transient monitors like Swift/BAT or Fermi/GBM as well as from serendipitous observations by INTEGRAL, this early emission would be invisible for astronomical distances.
Nevertheless, because there are still considerable uncertainties in these model parameters because classical novae have never been observed in the MeV $\gamma$-ray band, we propose directed and blind retrospective searches for nova outbursts in data archives.
Thanks to our parametrised spectro-temporal models, the sensitivity for such searches is greatly increased, and might reveal nearby outbursts that escaped UVOIR detection.

\section*{Acknowledgments}
S.C.L. acknowledges support by NASA grants HST-AR-15021.001-A and 80NSSC18K1017.
Thomas Siegert is supported by the German Research Foundation (DFG-Forschungsstipendium SI 2502/3-1) and acknowledges support by the Bundesministerium f\"ur Wirtschaft und Energie via the Deutsches Zentrum f\"ur Luft- und Raumfahrt (DLR) under contract number 50 OX 2201.

\section*{Data Availability}

The data underlying this article will be shared on reasonable request to the corresponding author.

\noindent This project is done with the use of software MESA \citep{Paxton2011,Paxton2013,Paxton2015,Paxton2018,Paxton2019} version 8118 and Python libraries: Matplotlib \citep{Matplotlib}, Pandas \citep{Pandas}, Numpy \citep{Numpy}.


\bibliographystyle{mnras}
\pagestyle{plain}
\bibliography{biblio,thomas}

\bsp	
\label{lastpage}
\end{document}